\documentclass[fleqn,usenatbib]{mnras}

\usepackage{newtxtext,newtxmath}

\usepackage[T1]{fontenc}

\DeclareRobustCommand{\VAN}[3]{#2}
\let\VANthebibliography\thebibliography
\def\thebibliography{\DeclareRobustCommand{\VAN}[3]{##3}\VANthebibliography}

\usepackage{graphicx}	
\usepackage{amsmath}	

\usepackage[normalem]{ulem}

\title[Feedback-Free Star Formation]{Feedback-Free Star Formation in Clusters within a Galaxy Simulated at High Resolution in Cosmic Dawn}

\author[H-Z. Chen et al.]{
Hou-Zun Chen,$^{1}$\thanks{E-mail: \href{mailto:chenhz_zju@zju.edu.cn}{chenhz\underline{ }zju@zju.edu.cn}}
Zhaozhou Li,$^{2,3,4}$\thanks{E-mail: \href{mailto:zhaozhou.li@nju.edu.cn}{zhaozhou.li@nju.edu.cn}}
Avishai Dekel,$^{4,5}$\thanks{This paper is dedicated to the memory of Prof. Avishai Dekel.}
Zhiyuan Yao, $^4$
Nir Mandelker, $^4$
and Xi Kang $^{1,6,7}$\thanks{E-mail: \href{mailto:kangxi@zju.edu.cn}{kangxi@zju.edu.cn}}\\
$^{1}$Institute for Astronomy, the School of Physics, Zhejiang University, Hangzhou 310058, China\\
$^{2}$School of Astronomy and Space Science, Nanjing University, Nanjing, Jiangsu 210093, China\\
$^{3}$Key Laboratory of Modern Astronomy and Astrophysics, Nanjing University, Ministry of Education, Nanjing 210093, China\\
$^{4}$Centre for Astrophysics and Planetary Science, Racah Institute of Physics, The Hebrew University, Jerusalem 91904 Israel\\
$^{5}$SCIPP, University of California, Santa Cruz, CA 95064, USA\\
$^{6}$Center for Cosmology and Computational Astrophysics, Zhejiang University, Hangzhou 310058, China\\
$^{7}$Purple Mountain Observatory, 10 Yuan Hua Road, Nanjing 210034, China
}

\pubyear{\the\year{}}

\begin{document}
\label{firstpage}
\pagerange{\pageref{firstpage}--\pageref{lastpage}}
\maketitle

\begin{abstract}
We perform a cosmological zoom-in simulation of a massive galaxy ($M_s\sim10^{10}\rm M_\odot$ at $z\sim10$) using the \texttt{GIZMO} code. By employing $\leq 3\rm pc$ resolution and a $3.4\rm Myr$ supernova feedback delay, we capture the feedback-free starbursts (FFB) in clusters. The simulation reproduces FFB model predictions and super-bright galaxies observed by JWST. At $z\sim10$, cold streams feed a compact galaxy ($R_{\rm e}\sim1\rm kpc$), with stellar and surface densities ($>10^5\rm cm^{-3}$, $>10^5\rm M_\odot pc^{-2}$) exceeding FFB thresholds. The global star-formation efficiency (SFE) is $\varepsilon_s\sim0.2\text{--}0.3$, associated with a fluctuating star-formation history. We identified over $10^5$ star clusters ($M_{\star}>10^{4.5}\rm M_\odot$) with a nearly scale-free mass distribution (${\rm d}N/{{\rm d}\log M}\propto M^{-1.06}$). Approximately 90\% of star formation occurs in clusters, which at a given time constitute $30\text{--}40\%$ of the total stellar mass. The star formation in most of the clusters of masses $<10^7\rm M_\odot$, occurs in bursts of $<3\rm Myr$ and a local SFE $\sim0.5\pm 0.2$. Cluster metallicities ($-2.01<\log (Z/Z_\odot)<-0.45$) indicate rapid baryon recycling. Feedback-driven outflows exhibit typical temperature of $10^7\rm K$ and typical velocities of $\sim 2000\rm km\ s^{-1}$. In the highly dynamic central $1\rm kpc$, clusters undergo rapid orbital decay and merge to assemble the oblate nuclear stellar cluster. Cluster shapes range from oblate to prolate, with a triaxial median. These clusters are consistent with JWST observations, and a fraction of them may survive to yield the globular clusters (GCs) at low redshifts.

\end{abstract}



\section{Introduction}\label{sec:1}

For the first time, observational results from the James Webb Space Telescope \citep[JWST, see][]{2022ApJ...940L..14N,2023Natur.616..266L} have paved the way for statistical analysis of brightest galaxies at $z=7\sim16$. Compared to predictions based on the widely accepted galaxy formation scenario within the $\Lambda$CDM framework, the observed one order-of-magnitude excess of UV-bright galaxies abundance \citep{2023ApJS..265....5H,2025ApJ...980..138H,2022ApJ...928...52F,2023ApJ...946L..13F,2023ApJ...951L...1P} has sparked numerous studies on possible features that may differ from galaxies at lower redshifts. Some actively pursued solutions in the recent literature include deviation from the standard cosmology \citep[e.g.][]{2021MNRAS.504..769K,2024MNRAS.533.3923S}, an ad hoc top-heavy initial mass function (IMF) of Pop-III stars \citep{2022MNRAS.509.4037Y}, redshift-dependence dust attenuation \citep{2023MNRAS.522.3986F}, bursty star formation histories (SFHs) of high-$z$ galaxies \citep[e.g.][]{2023ApJ...955L..35S,2023MNRAS.525.3254S}, and massive primordial black hole seeds \citep[e.g.][]{2022ApJ...937L..30L}.

Another promising explanation is the enhanced star formation efficiency (SFE) resulting from a less efficient stellar feedback mechanism at cosmic dawn ($z\sim10$). Based on the first-principle physical process, \cite{2023MNRAS.523.3201D} and \cite{2024A&A...690A.108L} (hereafter D23 and L24) show that the high densities and low metallicities of gas at $z\sim10$ naturally guaranteed a high SFE in dark matter haloes with mass $\sim10^{11} \rm M_{\odot}$. Feedback-free starbursts (FFB) occur when the freefall time is shorter than $\sim1\rm Myr$, with a corresponding gas density of $\rho\sim90\,\rm M_{\odot}pc^{-3}$, or equivalently $10^{3.5}\rm cm^{-3}$ (measured in unit of $\mu m_{\rm p}$, here $\mu=1.2$ is the mean molecular weight for neutral gas at $T\le10^4\rm K$). Specifically, based on the feedback mechanical energy injection rate computed by {\sc starburst99} \citep{1999ApJS..123....3L}, D23 shows that an instantaneous formed star cluster with $10^6\rm M_{\odot}$ will not develop sharp stellar feedback within $3\rm Myr$. When early stellar feedback is considered, a burst period of $\sim1\rm Myr$ is expected to be largely free of effective stellar-wind feedback and supernova feedback. D23 shows that this time window is robust across a range of metallicities and IMFs.

A natural consequence of the extreme gas densities required for the FFB mechanism is that star formation at cosmic dawn should be overwhelmingly clustered. Recent JWST observations \citep[such as][]{2024ApJ...963....9M} of compact, vigorous star-forming regions with effective radii smaller than $100\rm pc$ at $5<z<14$ lend further credence to the existence of these dense FFB environments. Theoretical models and recent high-resolution simulations also suggest that the intense, gas-rich environments of high-$z$ galaxies are ideal nurseries for bound massive star clusters. Furthermore, these compact stellar systems formed at $z\ge8$ have long been hypothesized as the progenitors of the metal-poor Globular Clusters (GCs) ubiquitous in the local Universe. For example, \cite{2016ApJ...823...52K} and \cite{2017ApJ...834...69L} demonstrated that the dense, gas-rich environments in cosmic dawn provide natural birthplaces for GC progenitors, characterized by rapid gas condensation and localized bursts of star formation within a few million years.


However, testing the evolutionary link between the extreme dense regions and clustered star formation requires understanding not only their initial formation efficiencies but also their subsequent dynamical fates. Nascent clusters face rapid internal feedback and intense external tidal fields; they may spiral into the galactic centre to build a nuclear star cluster (NSC), dissipate into the diffuse stellar halo, or survive to the present day. These possible evolutions can only be investigated in detail within the framework of modern hydrodynamical simulations. However, reproducing the FFB mechanism in hydro-simulations is quite challenging, as it requires modelling star formation in relatively massive environments with ultra-high spatial and mass resolution. Fortunately, the zoom-in technique allows for a significant increase in simulation resolution without a substantial increase in computational cost. 
In this work, we introduce a cosmological zoom-in hydro-simulation with ultra-high numerical resolution. The mass of gas particles within the finest region is $m_{\rm gas}=1125\mathrm{M_{\odot}}$, and the corresponding gravitational softening length is $\epsilon_{\rm gas}=3.0\mathrm{pc}$. The star cluster resolvable in our simulation has a mass of approximately $10^{4.5}\mathrm{M_{\odot}}$ (at least 30 star particles). 

This paper is organized as follows: In Section \ref{sec:2}, we provide an overview of the details of our simulation. Sections \ref{sec:3} to Section \ref{sec:7} present our main findings. In Section \ref{sec:3}, we present the overall properties of the simulated central galaxy, including its morphology, evolutionary history, star formation efficiency, and profiles. In Section \ref{sec:4} we explore the origin of the enhanced star formation, focusing on the statistical properties of the star clusters and gas clouds formed in our simulation. In Section \ref{sec:5} we study the star formation histories and metallicities of individual star clusters. Section \ref{sec:6} focuses on the dynamical evolution of star clusters. Section \ref{sec:7} analyse the cluster shapes. In Section \ref{sec:8} we briefly compare the properties of star clusters with local GCs. Conclusions are presented in Section \ref{sec:9}.

Throughout this paper, length scales are expressed in physical units, unless otherwise specified (e.g., ckpc denotes comoving units).

\section{Simulation and Methods}
\label{sec:2}

\subsection{Simulation}\label{sec:2.1}

We begin by selecting a high-$z$ halo from a cosmological $N$-body simulation with a periodic box length of $500h^{-1}\mathrm{Mpc}$ and $512^3$ dark matter particles, corresponding to a mass resolution of $8.14\times10^{10}h^{-1}\rm{M_{\odot}}$. At $z=7.7$, we identify one of the most massive haloes in the volume, with a virial mass of $M_{\rm vir}=1.8\times10^{12}h^{-1}{\rm M_\odot}$, where the virial mass is defined as the mass enclosed within a spherical overdensity of 200 times the cosmic critical density. This halo is then selected for high-resolution re-simulation using the publicly available code \texttt{GIZMO} \citep{2015MNRAS.450...53H}, a parallel hydrodynamic simulation framework that supports multiple solvers and physics modules. We perform a zoom-in re-simulation of this massive galaxy from $z=99$ (the initial condition redshift) to $z=8$, incorporating baryonic physics and feedback prescriptions. The simulation outputs 110 snapshots at 5Myr intervals, spanning from 80Myr (corresponding to $z\sim35$, just before the first star particle forms) to 630Myr. To better resolve the dynamical evolution of gas clouds and star clusters, we generated 75 additional snapshots between 445Myr ($z=10.46$) and 460Myr ($z=10.20$) with a finer output interval of 0.2Myr. This allows us to capture the rapid formation and evolution of clusters in detail (see Section \ref{sec:5.4} and Section \ref{sec:6}). 


The cosmological parameters adopted in this work are from \cite{2014A&A...571A..16P}, specifically, $\Omega_{\rm m}=0.3175$, $\Omega_{\Lambda}=0.6825$, $\Omega_{\rm b}=0.049$, $H_0=67.1 \mathrm{km\,s^{-1}Mpc^{-1}}$, and $\sigma_8=0.8344$. The initial conditions for both the original $N$-body simulation and the zoom-in simulation are generated using the \texttt{MUSIC} code \citep{2011MNRAS.415.2101H}. The finest region has a comoving volume of $1.85\times0.93\times1.85 h^{-3}\rm{cMpc^3}$ in the initial condition, which contains $2.4\times10^8$ finest particles for both dark matter and gas. As a result, the mass resolution is $m_{\rm gas}=1125\mathrm{M_{\odot}}$ and $m_{\rm dm}=6107\mathrm{M_{\odot}}$. Following \cite{2005MNRAS.364.1105S}, \texttt{GIZMO} code uses
\begin{equation}
    \epsilon=\min\left(\epsilon_{\rm com}/(1+z),\ \epsilon_{\rm phys}^{\rm max}\right),
\end{equation}
as the physical softening. In our simulation, we set $\epsilon_{\rm com}=30\mathrm{pc}$ and $\epsilon_{\rm phys}^{\rm max}=3\mathrm{pc}$ for gas particles, $\epsilon_{\rm com}=80\mathrm{pc}$ and $\epsilon_{\rm phys}^{\rm max}=8\mathrm{pc}$ for dark matter particles. So the gravitational softening was kept fixed in comoving units until $z=9$ and in proper units thereafter.

We employ the meshless finite-mass (MFM) method, which is also the default solver of the FIRE project \citep{2018MNRAS.480..800H}, to solve the hydrodynamics. The cooling and heating processes of primordial gas (i.e. H and He) are from an alternative nonequilibrium chemical (ion + atomic + molecular) network, namely the CHIMES module, mainly developed by \cite{2014MNRAS.440.3349R,2014MNRAS.442.2780R}. For the metal part, we trace nine elements separately, that is, C, N, O, Ne, Mg, Si, S, Ca and Fe \citep[for details refer to][]{2014MNRAS.445..581H}. Cold gas particle turns into star particle stochastically when it is locally self-gravitating, self-shielding, Jeans unstable, and above a certain volume density threshold (specifically, we set $n_{\rm th}=5000\rm cm^{-3}$, the reason will be discussed below). Once a star particle is generated, it will inherit the metallicity and mass from the progenitor gas particle. The supernova (SN) rate is inferred separately from Type Ia and Type II by assuming different event rates \citep[refer to Appendix A of FIRE-2 paper, i.e.][and also the discussion in Section \ref{sec:2.2}]{2018MNRAS.480..800H}. We then employ the mechanical feedback model to tackle the subsequent SN feedback events from each star particle. For detailed methodology on this feedback algorithm, please refer to \cite{2014MNRAS.445..581H,2018MNRAS.477.1578H}. We identified star clusters and gas clouds using the standard friend-of-friend (FoF) algorithm \citep{1985ApJ...292..371D}. Lastly, we used the \texttt{PYNBODY} package \citep{2013ascl.soft05002P} to analyse our simulation data. Note that except for the specific density threshold for star formation, which is related to the mass resolution, this set of recipes is also applied in the NIHAO-RiNG project \citep{2024ApJ...977..233C}, in which we investigated the impact of different feedback recipes on the properties of the circumgalactic medium in Milky Way-like disk galaxies at $z\sim0$. 

\subsection{Numerical Methods Specific for FFB Scenario}\label{sec:2.2} 

\begin{figure}
    \includegraphics[width=\linewidth]{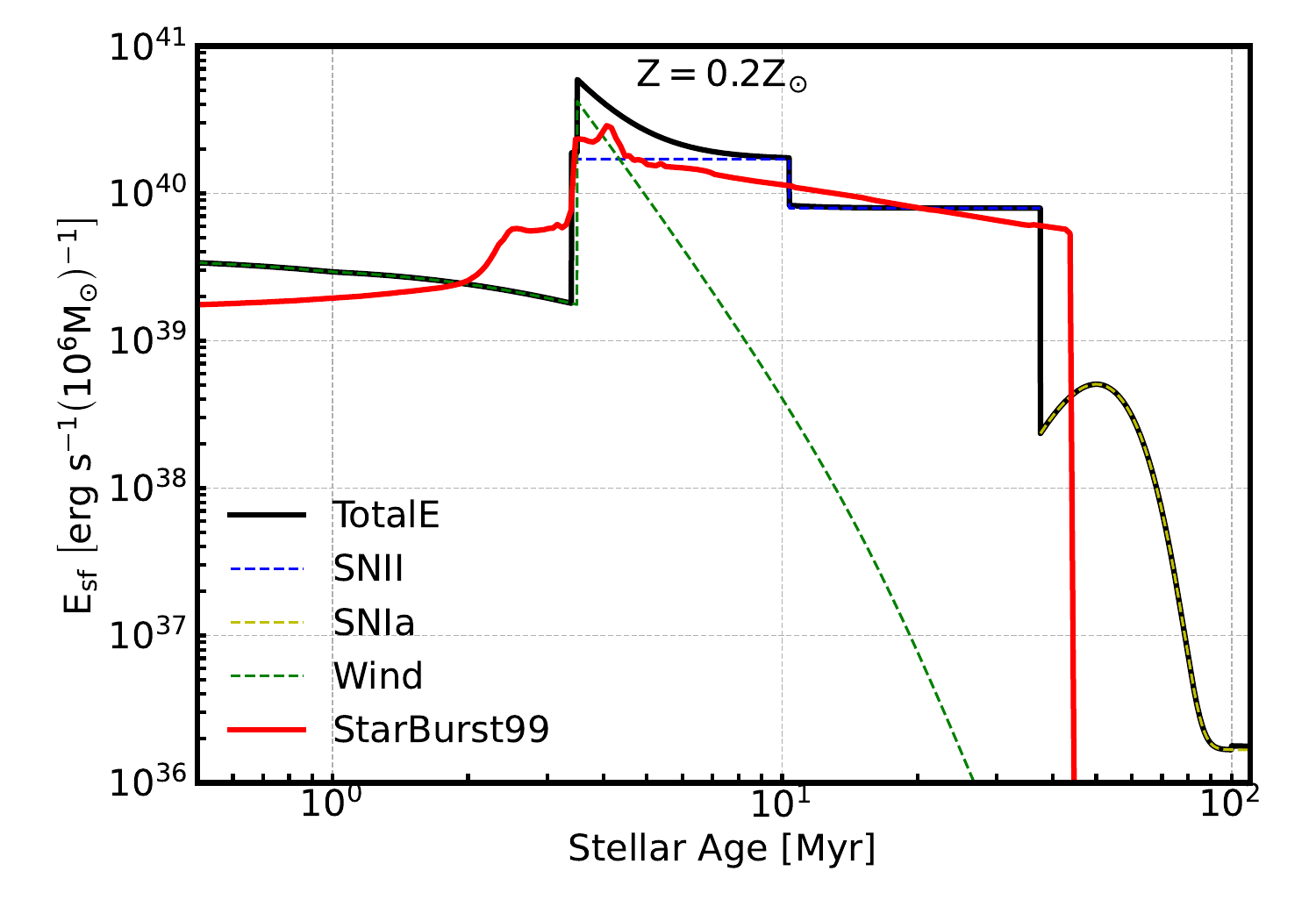}
    \caption{The stellar feedback energy injection rate implemented in \texttt{GIZMO} for an instantaneous starburst in a star cluster of $10^6\rm M_{\odot}$ and 0.2 solar metallicity. Dashed curves with various colours show different stellar feedback source, and black solid line shows the total feedback energy. As comparison, the red solid line gives the feedback energy injection rate as computed by {\sc starburst99} assuming Kroupa IMF \citep{2001MNRAS.322..231K} and 0.2 solar metallicity.}
    \label{fig:stellarfeedback}
\end{figure}

As introduced in Section \ref{sec:1}, the FFB scenario occurs when the freefall time in the star-forming region is shorter than the feedback timescale. Figure \ref{fig:stellarfeedback} shows the stellar feedback energy injection rate implemented in \texttt{GIZMO} for an instantaneous starburst in a star cluster of $10^6\rm M_\odot$ \citep[the piecewise fitting functions of various feedback source are from Appendix A of][]{2018MNRAS.480..800H}. As seen, the onset of stellar feedback from both Type Ia and Type II SN does not occur until at least $3.4\rm Myr$ after the formation of star particle; prior to this point, the stellar wind feedback is comparatively weak, consistent with the delayed feedback predicted by {\sc starburst99} (the red curve, also see Figure 1 in D23 for other metallicity and IMF cases). 

Another crucial condition for the FFB scenario is the gas number density, which is related to the free-fall time $t_{\rm ff}$ by
\begin{equation}
    t_{\rm ff}=\sqrt{\frac{3\pi}{32G\rho_{\rm gas}}}\simeq0.84n^{-0.5}_{3.5}\mathrm{Myr}\;,
\end{equation}
where $n_{\rm gas}=10^{3.5}n_{3.5}\,\mathrm{cm^{-3}}$. To properly resolve the FFB scenario, the star formation threshold $\rho_{\rm th}$ must exceed this characteristic density.
On the other hand, the maximum gas density at which gravitational instabilities can be properly resolved determines the numerical upper limit for this threshold:
\begin{equation}
    \rho_{\rm th}\le \rho_{\rm gas,max}=N_{\rm ngb}m_{\rm gas}\epsilon^{-3}_{\rm gas}\simeq10^4\mathrm{cm^{-3}},
\end{equation}
where $N_{\rm ngb}=32$ is the number of neighbours used to estimate the gas density, $\epsilon_{\rm gas}\sim3\rm pc$ is the maximum physical softening length of gas. 
Therefore, an appropriate density threshold for star formation must be established within the range $\left[10^{3.5},\,10^4\right]\rm cm^{-3}$, and we adopt $\rho_{\rm th}=5000\rm cm^{-3}$ hereafter. The above two specific configurations make \texttt{GIZMO} a suitable code for simulating the FFB scenario at high redshifts.

\section{Galaxy Properties}\label{sec:3}
\subsection{Overall Morphology}\label{sec:3.1}

\begin{figure*}
    \includegraphics[width=0.49\linewidth]{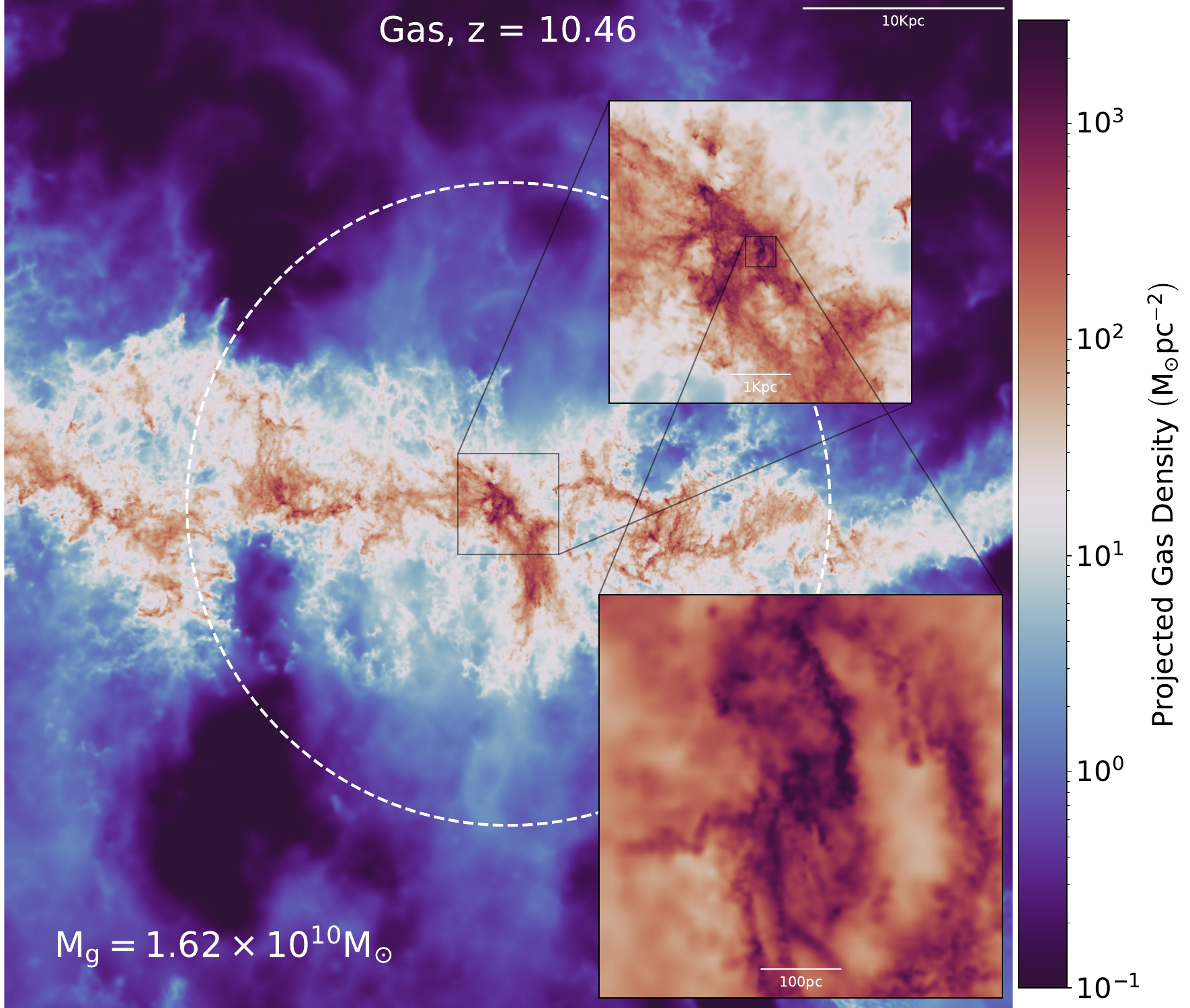}
    \includegraphics[width=0.49\linewidth]{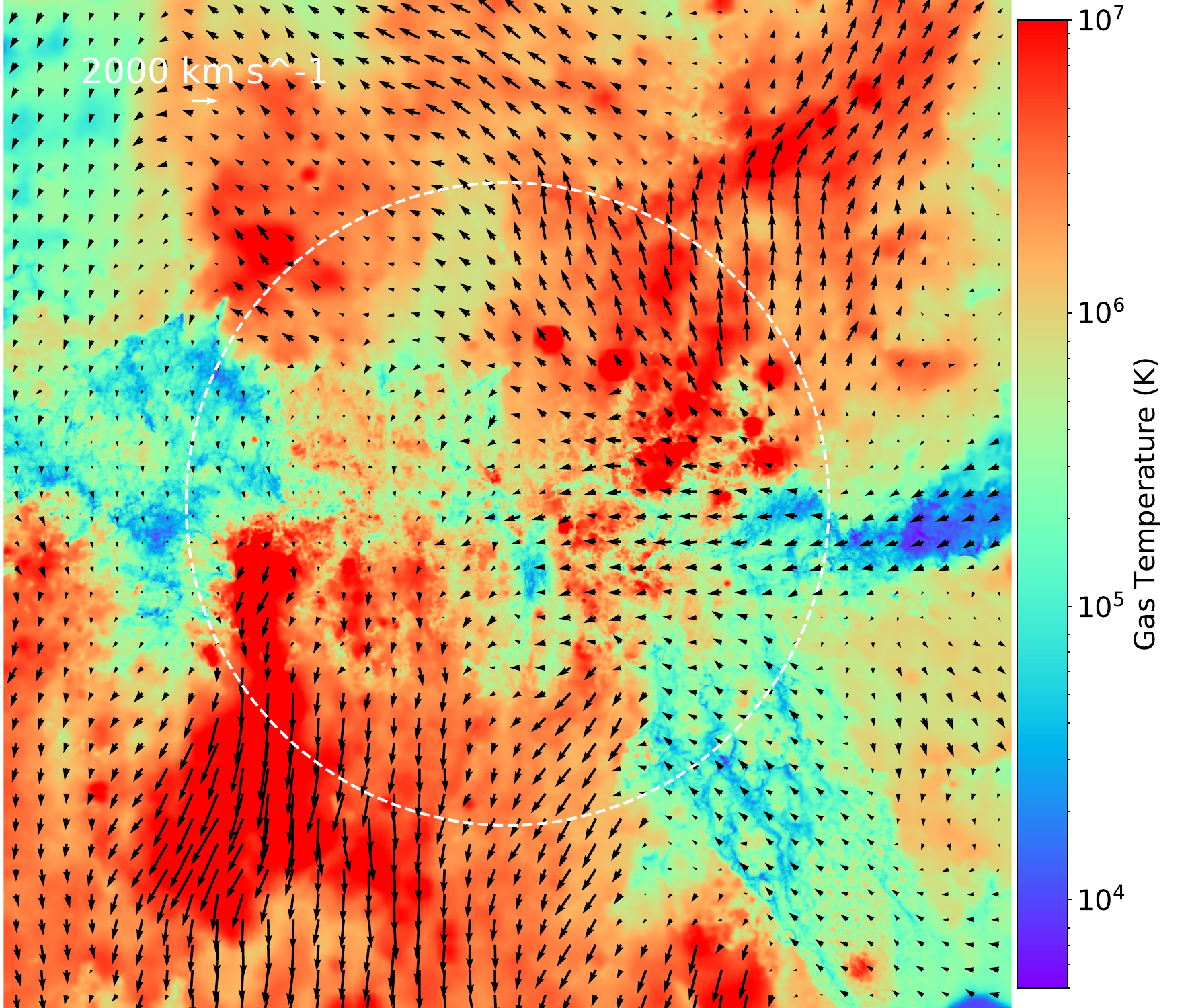}
	\includegraphics[width=0.49\linewidth]{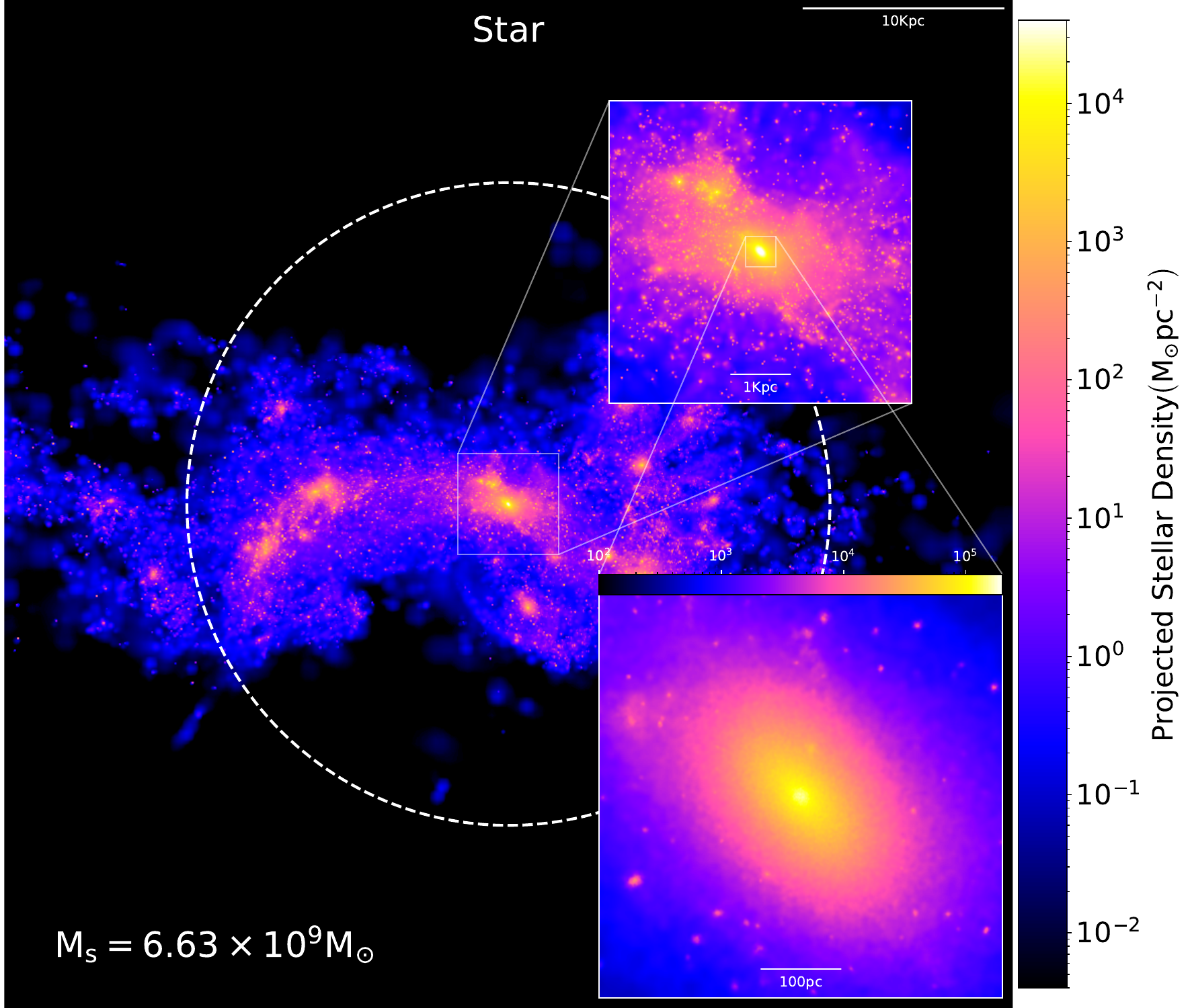}
    \includegraphics[width=0.49\linewidth]{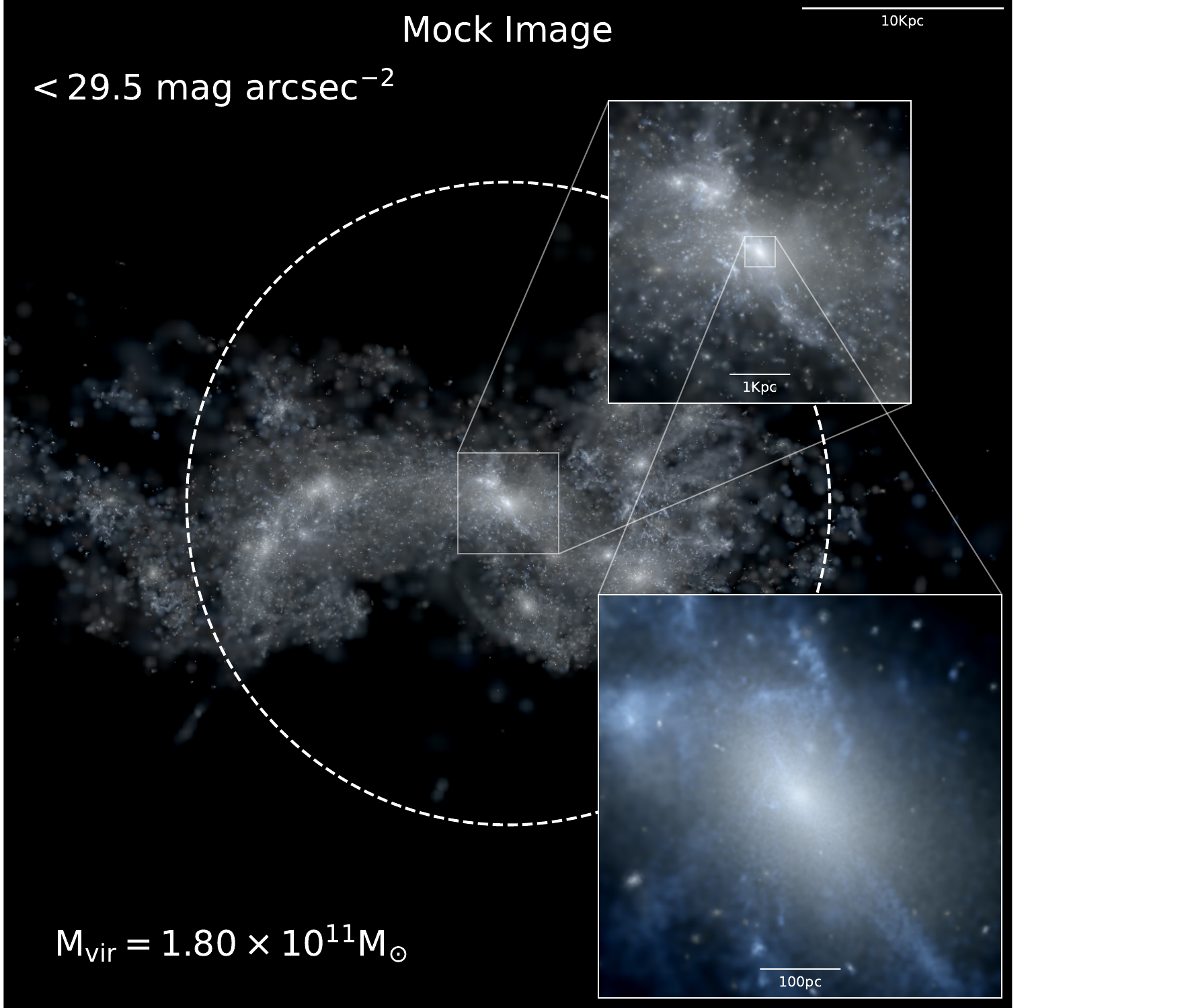}
    \caption{A $50\rm kpc\times50kpc\times10kpc$ slice of the simulated massive galaxy at $z=10.46$. The left panels show the projected gas density (top) and stellar density (bottom). The upper‐right panel displays the line‐of‐sight, density‐weighted temperature map overlaid with gas velocity vectors. The lower‐right panel presents a mock optical image in the \textit{i}, \textit{v} and \textit{u} bands, with a surface‐brightness limit of $29.5\rm mag\,arcsec^{-2}$ (comparable to JWST data). The dashed circle indicates the virial radius. In the second, third, and fourth quadrants, we annotate the total gas mass, stellar mass, and virial mass enclosed within the virial radius, respectively. Each of these panels also includes two inset plots showing a cascaded zoom‐in view. Note that the colour bar for the lower‐right inset in the stellar density panel has been rescaled to accommodate its large density range.}
    \label{fig:zoom_ffb}
\end{figure*}

In this section, we present the overall morphology of the simulated massive galaxy. Figure \ref{fig:zoom_ffb} shows a $50\rm kpc\times50kpc\times10kpc$ slice of this system at $z=10.46$. The total mass of the halo (including baryon and dark matter) reaches $1.8\times10^{11}\rm M_{\odot}$ and the stellar mass exceeds $6\times10^{9}\rm M_{\odot}$. Firstly, the two gas panels show that filamentary cold inflows with temperatures of $10^{4-5}\rm K$ penetrate from outside the virial radius towards the halo centre, indicating that the entire system is in the cold-flow regime \citep{2006MNRAS.368....2D}. In this regime, the radiative cooling time of gas is much shorter than the halo crossing time. The filamentary morphology of the cold inflow is important because it delivers gas supplies efficiently and is largely resistant to feedback outflows. As shown in the gas temperature map, feedback-driven outflows exhibit typical temperatures of $10^7\rm K$ and typical velocities of $\sim2000\rm km/s$, consistent with the predictions in L24 (Equation 22), yet they fail to disrupt the accretion, venting primarily into the diffuse, low-density regions between the filaments. Consequently, the cold inflows from the horizontal filaments converge and collide within the central $1\rm kpc$ (physical scale; hereafter implied), driving the projected gas density up to ${\sim}10^3\rm M_{\odot}\mathrm{pc}^{-2}$. 

Secondly, the projected stellar density map reveals a compact NSC at the halo centre with a physical scale of $\sim100\rm pc$ (lower right inset in the lower left panel). This nucleus reaches an extraordinary projected mass density exceeding $10^{5}\rm M_{\odot}pc^{-2}$ – two orders of magnitude higher than the gas density in the same region. This projected density is consistent with the lensed star clusters reported by  \cite{2024Natur.632..513A} in the JWST observations of the Cosmic Gems arc at $z\sim10.2$. The spatial size of this ultra-dense core, evident in both the density map and mock images, is roughly consistent with recent JWST observations of vigorous star‑forming regions in high-$z$ galaxies \citep{2024ApJ...963....9M}. Interestingly, as noted by \cite{2010MNRAS.401L..19H}, the stellar surface density of $10^5\rm M_\odot pc^{-2}$ aligns with a wide array of observed dense stellar systems: globular clusters, super-star clusters, nuclear clusters of dwarf and late-type galaxies, young massive clusters, ultra-compact dwarfs, compact ellipticals, galactic bulges, nearby and high-$z$ early-type galaxies, spanning over 7 orders of magnitude in total stellar mass.

Thirdly, the inset plots of the stellar panel and the mock image panel reveal a large population of compact star clusters around the NSC. Our analysis supports the idea that these compact star clusters are formed via the FFB scenario, in which stars formed originate in localized, high-density regions capable of self-shielding against feedback from earlier star clusters. A detailed discussion on their formation and evolution will be given in Section \ref{sec:5} and Section \ref{sec:6}.

\subsection{Overall Evolution}\label{sec:3.2}

\begin{figure}
    \includegraphics[width=\linewidth]{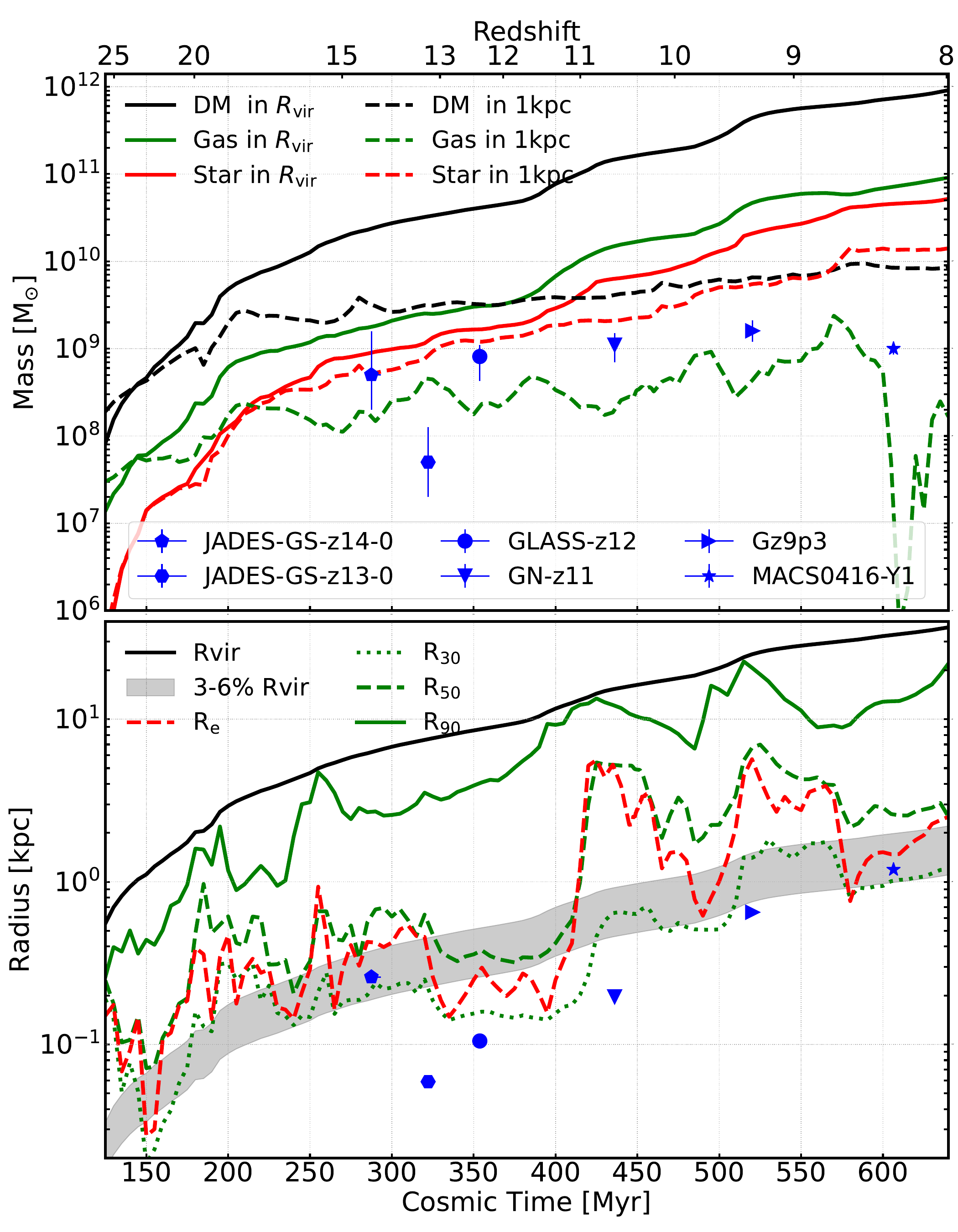}
    \caption{Basic properties of the simulated massive galaxy as function of cosmic time. The panels arranged from top to bottom illustrate: (top) the mass accretion histories of different components. Solid curves represent dark matter, gas, and stellar mass growth within $R_{\rm vir}$, dashed curves show corresponding growth within the $1\rm kpc$ range; (bottom) the evolution of five characteristic radii: the virial radius ($R_{\rm vir}$ in black curve), the half light radius ($R_{\rm e}$ in red dashed curve), and the radii enclosing 30\% , 50\% and 90\% of the total stellar mass ($R_{30}$ in dotted curve, $R_{50}$ in green dashed curve and $R_{90}$ in solid green curve), respectively. The shaded band represents the $3-6\%R_{\rm vir}$ range. As comparison, 6 high-$z$ galaxies are listed. The upper panel shows their total stellar mass, while the bottom panel shows their effective radius. These results are from: JADES-GS-z14-0 \citep{2025NatAs...9..729H}, JADES-GS-z13-0 \citep{2023NatAs...7..611R,2024ApJ...976..160H}, GLASS-z12 \citep{2024ApJ...975..245C}, GN-z11 \citep{2023A&A...677A..88B,2023ApJ...952...74T}, Gz9p3 \citep{2024NatAs...8..657B}, MACS0416-Y1 \citep{2024ApJ...977L..36H}.}
    \label{fig:infos}
\end{figure}

Figure \ref{fig:infos} presents the mass and size evolution of the simulated massive galaxy during $z=25\sim8$. For comparison, we selected 6 confirmed high-$z$ galaxies at $z>8$, and marked their redshifts, stellar mass, and half light radius in the two panels. As shown, our simulated sample is comparable in mass and size to the observed high-$z$ galaxies. The top panel traces the mass growth histories of different components in different ranges. While the total masses of dark matter, gas and star within $R_{\rm vir}$ follow similar growth trends, the central $1\rm kpc$ region exhibits different evolutionary pathways. Between $z=18$ and $z=8$, the dark matter mass within this region increases slowly, and the gas mass fluctuates between $\left[10^8,\,10^9\right]\rm M_{\odot}$ (the bump above $10^9\rm M_{\odot}$ at $z\simeq8.7$ is driven by a major merger event, which eventually cleans up all the central gas due to subsequent starbursts). In contrast, the stellar mass grows significantly from $10^8\rm M_{\odot}$ to $10^{10}\rm M_{\odot}$ in the same period, leading to the dominance of the stellar component within the central $1\rm kpc$. 

The lower panel traces five characteristic radii of the system: the virial radius ($R_{\rm vir}$), the half–light radius ($R_{\rm e}$), the half-stellar mass radius ($R_{50}$) and the radii enclosing 30\% and 90\% of the total stellar mass ($R_{30}$, $R_{90}$). Up to $z=11$, both $R_{50}$ and $R_{\rm e}$ remain compact ($<1\rm kpc$). The sharp increases in these radii at $z\simeq11$ and $z\simeq9.5$ correspond to two major merger events, also indicated by the intersection of $R_{\rm vir}$ and $R_{90}$. The divergent evolution of $R_{50}$ and $R_{90}$ demonstrates that the stellar distribution extends significantly beyond the half‑mass radius. Meanwhile, the growth of $R_{30}$ and $R_{\rm e}$ scales roughly as $0.03\sim0.06R_{\rm vir}$. Note that recently \cite{2025arXiv250505442S} found that a ratio $R_{\rm e}/R_{\rm vir}\sim5\%$ is needed for high-$z$ galaxies, which is roughly consistent with our findings.  

\subsection{Star Formation Efficiency}\label{sec:3.3}

\begin{figure}
    \includegraphics[width=\linewidth]{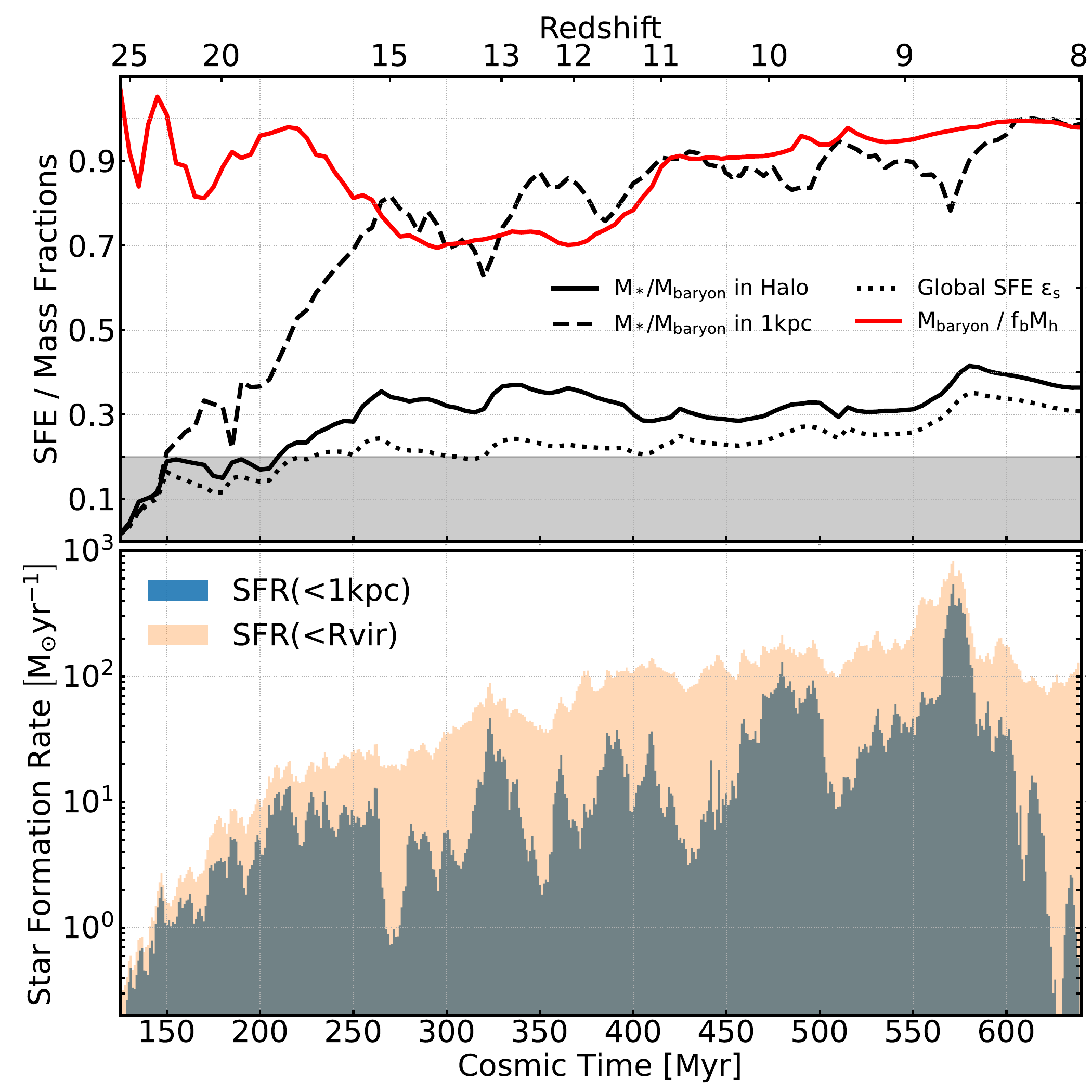}
    \caption{The evolution of star formation efficiency, stellar mass fraction and baryon fraction. The solid and dashed curve represent stellar mass fraction (see definition in text) within $R_{\rm vir}$ and $1\rm kpc$ range, respectively. The dotted curve shows the evolution of global star formation efficiency $\varepsilon_s$. The shaded region highlights $\rm\varepsilon_s<0.2$; (bottom) the star formation histories within $R_{\rm vir}$ (orange histogram) and $1\rm kpc$ (blue histogram).}
    \label{fig:infos2}
\end{figure}

The top panel of Figure \ref{fig:infos2} tracks the evolution of the SFE and baryon mass fractions during $z=25\text{--}8$. We define the global SFE, $\varepsilon_s$, and the stellar mass fraction within a given radius, $f_s(r)$, as
\begin{equation}
    \varepsilon_s=\frac{M_s}{\bar{f_b}M_{\rm vir}}, \quad f_s(r)=\frac{M_{s}(<r)}{M_{\rm baryon}(<r)},
\end{equation}
where $\bar{f_b}=0.16$ is the universal baryon fraction. Since $z\sim18$, $\varepsilon_s$ fluctuates between $20\text{--}35\%$ and remains safely above the $\varepsilon_s=0.2$ threshold implied by JWST UV luminosity functions at $z\sim10\text{--}14$ \citep{2025ApJ...980..138H}. Meanwhile, the stellar mass fraction within the virial radius, $f_s(R_{\rm vir})$, tracks slightly higher at $30\text{--}40\%$; this is because the global baryon fraction (red curve) is suppressed to roughly $0.7\bar{f_b}$ throughout $z=18\text{--}11$ due to the feedback-driven outflows discussed in Figure \ref{fig:zoom_ffb}. In distinct contrast to the halo scale, the central $1\rm\,kpc$ becomes strongly stellar-dominated, with $f_s(1\rm\,kpc)$ rising steadily from $30\%$ to $\sim 90\%$.

It is informative to compare our global SFE with another two recent high-resolution hydro-simulations done by \cite{2025MNRAS.540.3350A} and \cite{2025ApJ...981L..28M}, who also investigated galaxy formation in a highly overdense region at cosmic dawn using the \texttt{RAMSES} code and \texttt{GASOLINE2} code, respectively. The SNe-only run from \cite{2025MNRAS.540.3350A} reaches a $\varepsilon_s \simeq 0.34$ at $z\sim9$. The halo mass of primary galaxy simulated by \cite{2025ApJ...981L..28M} exceeding $10^{12}\rm\ M_{\odot}$ and the stellar mass reaches $M_s\left(<2\rm kpc\right)=8\times10^{10}\rm M_{\odot}$ at $z\simeq7.6$, which also implies a high underlying global SFE comparable to this work.

The bottom panel shows the SFHs of the total halo ($<R_{\rm vir}$) and the inner $1\rm kpc$ region. The logarithmic scale highlights that the inner SFH is significantly more bursty than the total SFH of the halo. Specifically, the merger events at $z\sim10$ and $z\sim8.8$ drive two pronounced peaks. At $z\sim10$, the SFH reaches $\sim100\rm M_{\odot}yr^{-1}$, in reasonable agreement with the $\sim65\rm M_{\odot}yr^{-1}$ predicted by D23. 

An analysis of the detrended SFH within $1\rm kpc$ using the autocorrelation function (ACF) reveals a bursty star formation history with a primary coherence timescale (at $1/e$) of $\sim10-20\rm Myr$ (range for uncertainties from the detrending window size and the choice of linear versus logarithmic SFR), which is consistent with the burst duty cycles predicted by D23 and L24. Additionally, a secondary ACF peak at $\sim80-90\rm Myr$ reflects a longer-term modulation, potentially driven by the typical separation between merger events or low-frequency variations in the global accretion rate.

\subsection{Density Profiles}\label{sec:3.4}

\begin{figure}
    \includegraphics[width=\linewidth]{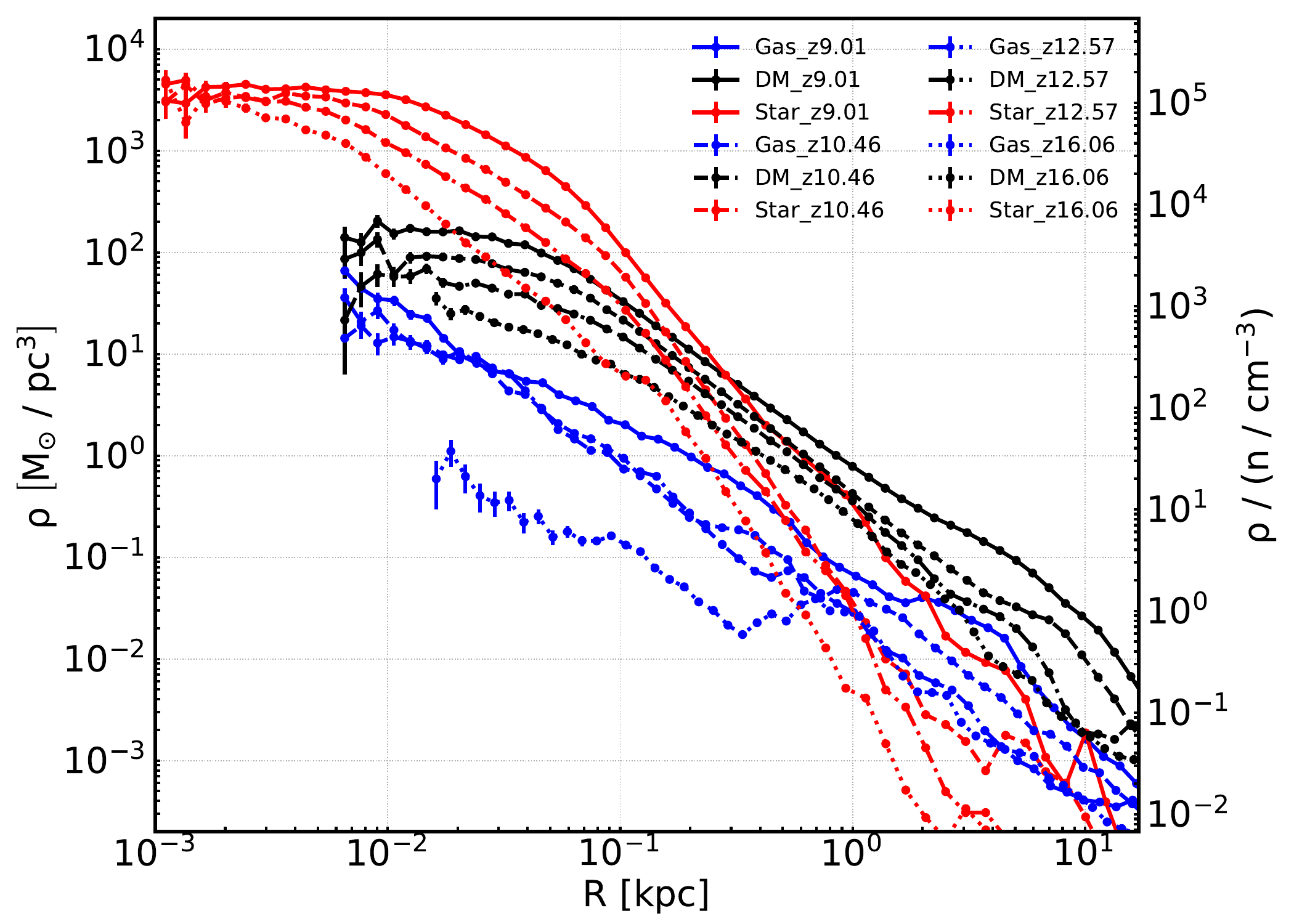}
    \caption{Spherically averaged radial profiles of the stellar (red), gas (blue), and dark matter (black) components of the central galaxy. The dotted, dash-dotted, dashed, and solid lines correspond to redshifts $z=16.06$, 12.57, 10.46, 9.01, respectively.}
    \label{fig:prof}
\end{figure}

Figure \ref{fig:prof} shows the three-dimensional density profiles of the dark matter, stellar, and gaseous components at four representative snapshots spanning $z=16$ to $z=9$. Four principal findings emerge:
\begin{enumerate}
    \item The central stellar density remains constant at $\sim10^{5}\rm cm^{-3}$ (equivalently $10^{3.5}\rm M_{\odot}pc^{-3}$) throughout the redshift range, which is comparable to the central stellar density of non-core-collapsed globular clusters in our Milky Way and other nearby galaxies \citep{1996AJ....112.1487H,2005ApJS..161..304M}.
    \item With the exception of the earliest snapshot ($z=16.06$), the maximum central gas density reaches $\sim10^3\rm cm^{-3}$.
    \item The stellar component dominates the mass distribution within the inner region, which expands from approximately $100\rm pc$ to $200\rm pc$ over the period studied. Beyond $1\rm kpc$, the gaseous component becomes the main contributor of baryon mass.
    \item Both the stellar and dark matter density profiles increase steadily with time at all radii, showing that the central galaxy follows the inside-out evolution.
\end{enumerate}

\section{Population of Star Clusters and Gas Clouds}\label{sec:4}
\subsection{Cluster Identification}\label{sec:4.1}

\begin{figure*}
    \includegraphics[width=0.48\linewidth]{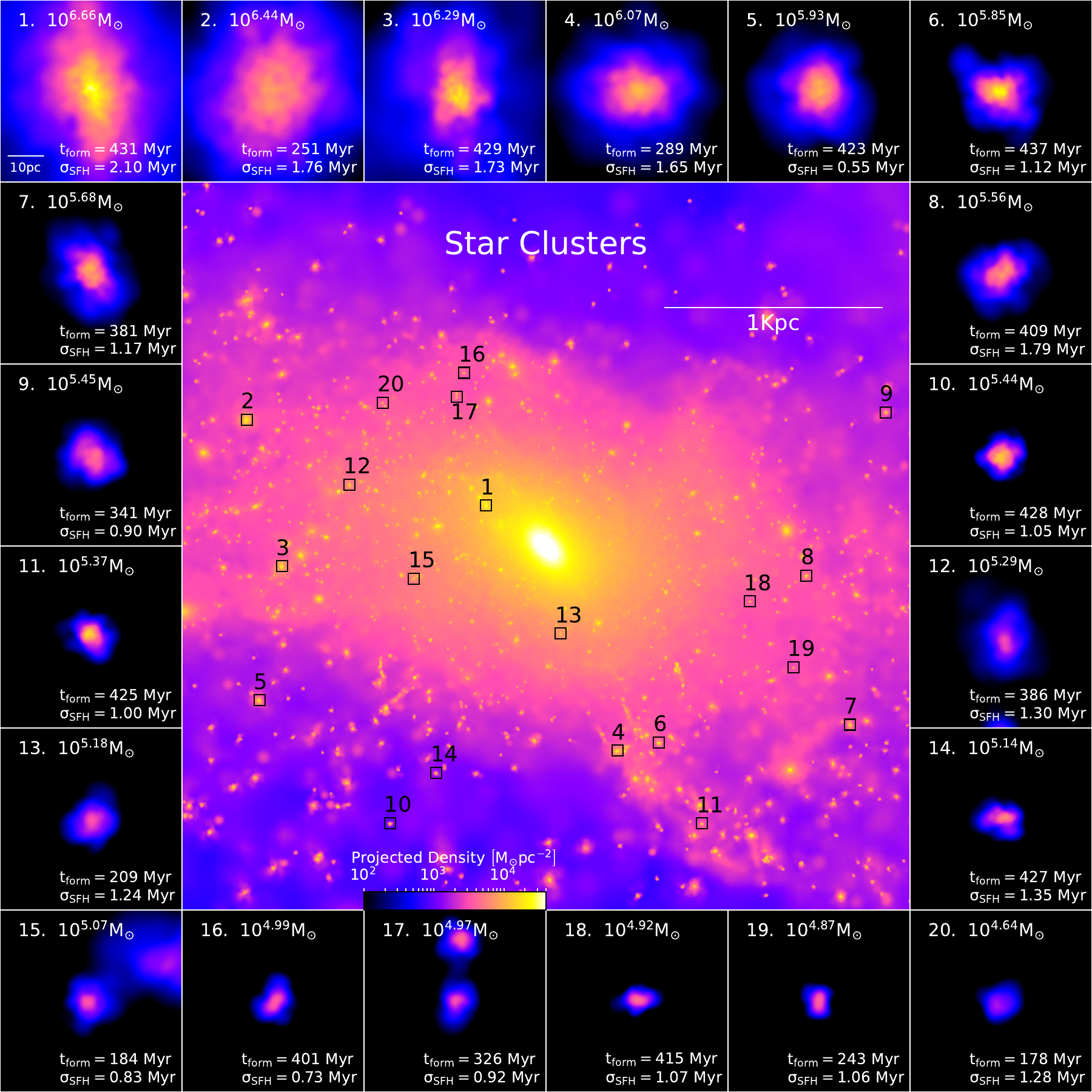}
    \includegraphics[width=0.48\linewidth]{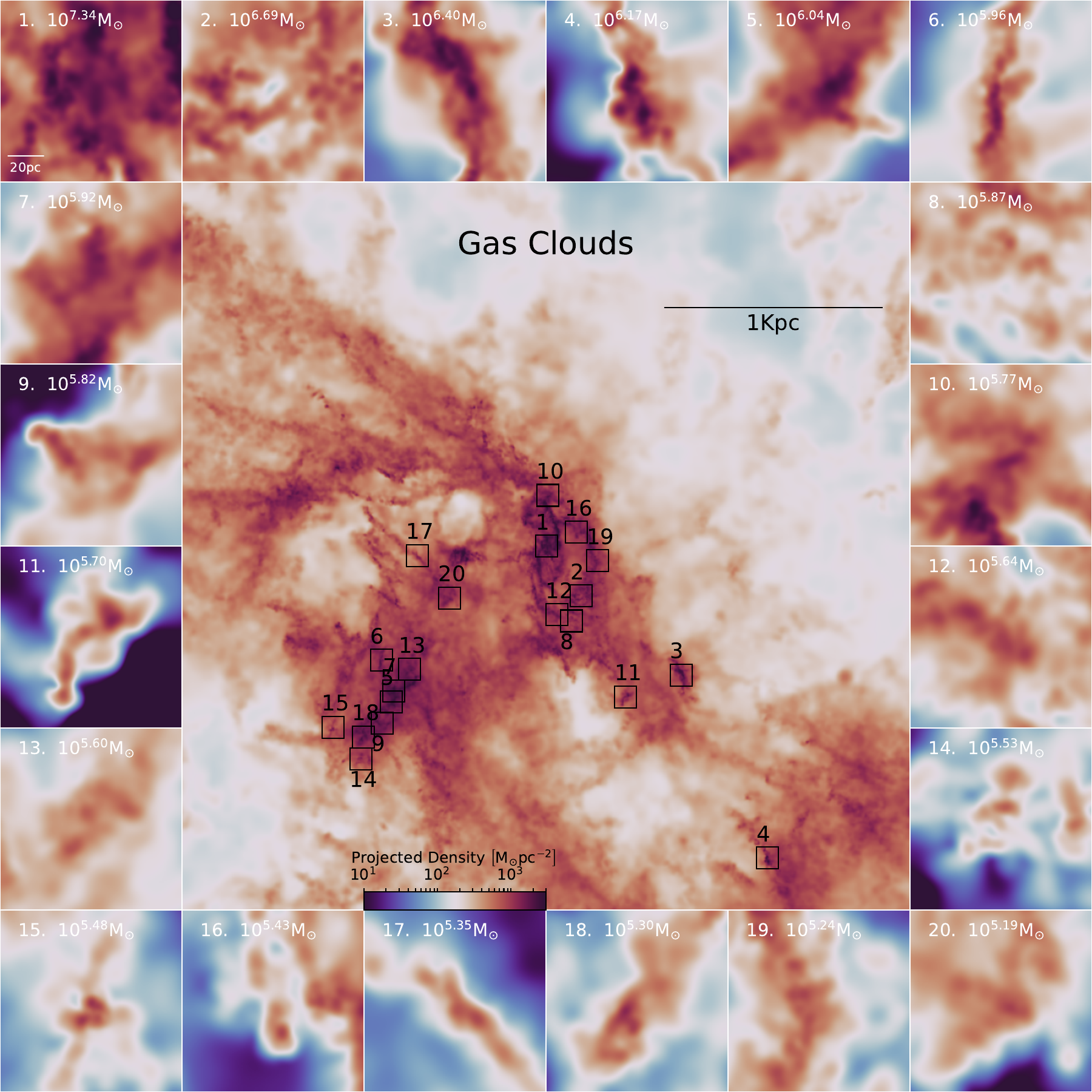}
    \caption{Morphology and spatial distribution of 20 randomly selected star clusters (left) and dense gas clouds (right) identified using the FoF algorithm in our simulation. Cluster/Cloud masses are indicated at the top of each panel. For star clusters, both the formation time ($t_{\rm form}$) and the standard deviation of their star formation histories ($\sigma_{\rm SFH}$) are provided (see text for details). The colour scheme of two main panels are identical to that present in Figure \ref{fig:zoom_ffb}, while the colour bars shows the projected densities.}
    \label{fig:sgclusters}
\end{figure*}

As revealed in Figure \ref{fig:zoom_ffb}, numerous star clusters are distributed around the central galaxy, consistent with predictions that abundant clusters form prior to the epoch of reionization \citep[e.g.,][]{2018ApJ...861..148M,2024arXiv241102502C}. In the FFB scenario, the high gas densities in the ISM trigger violent gravitational instabilities that fragment into dense clouds; these clouds then rapidly convert into star clusters with high SFE on short timescales.

To identify these structures, we employ the standard Friends-of-Friends (FoF) algorithm \citep{1985ApJ...292..371D} with a fixed linking length of $b=5\rm\,pc$ to select star clusters and dense gas clouds from the stellar and gas particle distributions, respectively. Sensitivity tests indicate that the identified structures are robust against variations in linking length between $2\text{--}10\rm\,pc$. To ensure statistical reliability, we restrict our analysis to clusters containing at least 30 particles, corresponding to a lower mass limit of $\sim 10^{4.5}\rm\,M_{\odot}$. Figure \ref{fig:sgclusters} illustrates the morphology of 20 randomly selected stellar (left) and gaseous (right) clusters surrounding the NSC at $z=10.46$. The star clusters typically exhibit compact, spheroidal morphologies, whereas the gas clouds are characterized by filamentary structures.

\subsection{Mass Fraction in Clusters}\label{sec:4.2}

\begin{figure}
    \includegraphics[width=\linewidth]{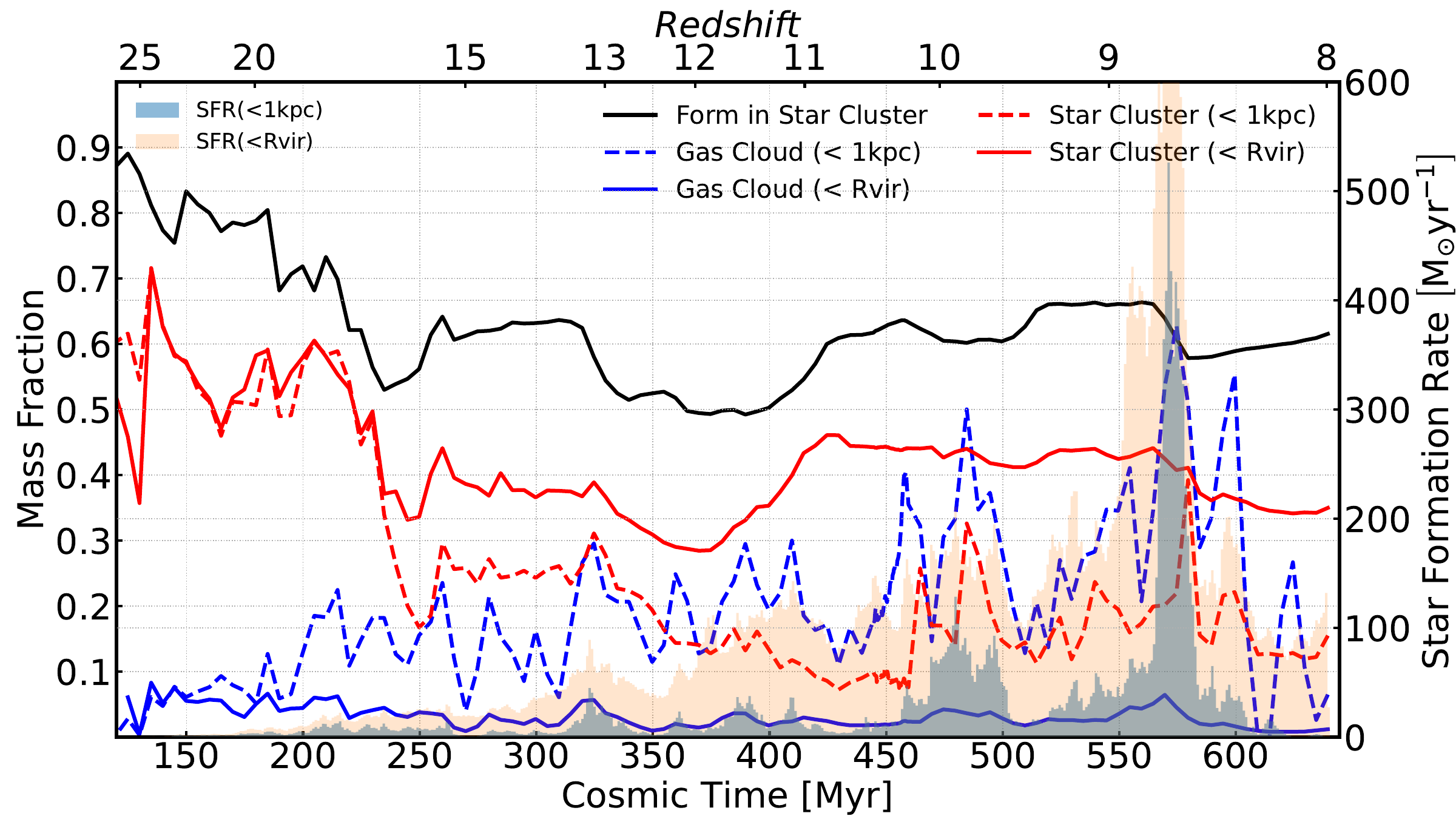}
    \caption{The mass fraction of gas cloud and star cluster within different radii. The orange and blue histogram show the star formation rates within $R_{\rm vir}$ and $1\rm kpc$ range, which is the same as the bottom panel of Figure \ref{fig:infos2} but in linear scale. The red and blue curves represent the mass fraction of star clusters and gas clouds in total stellar mass and gas mass (see definitions in text), respectively. The solid black curve represents the mass fraction of the star that within $R_{\rm vir}$ and formed in a FFB star cluster (also see definition in text). The solid and dashed curves correspond to the results within $R_{\rm vir}$ and the central $1\rm kpc$, respectively.}
    \label{fig:sfrfrac}
\end{figure}

Figure \ref{fig:sfrfrac} presents the evolution of the mass fractions for dense gas clouds and star clusters. We define these fractions as the total mass of identified objects (with $M \ge 10^{4.5}\rm\,M_{\odot}$, excluding the NSC) divided by the total gas or stellar mass within a specified radius ($R_{\rm vir}$ or the central $1\rm\,kpc$).

In the central $1\rm\,kpc$, the mass fraction of dense gas clouds fluctuates significantly between $10\text{--}50\%$, whereas on the halo scale, it remains consistently below $10\%$. Given that the total gas mass within $1\rm\,kpc$ varies between $10^8\text{--}10^9\rm\,M_{\odot}$ (Figure \ref{fig:infos}), the total mass in dense clouds within the core rarely exceeds $5\times10^8\rm\,M_{\odot}$. The stellar cluster mass fraction in the central region generally remains lower than the global fraction within $R_{\rm vir}$  and exhibits distinct evolutionary trends. While the halo-scale fraction remains relatively stable at $30\text{--}40\%$, the central fraction closely tracks the oscillations of the gas cloud fraction and the inner SFH. This correlation suggests that the burstiness of the inner SFH is directly driven by the rapid, episodic conversion of dense gas clouds into star clusters.

To distinguish between cluster formation and subsequent dynamical disruption, we compare the instantaneous bound fraction with the cluster birth fraction, defined as the mass fraction of stars that originally formed within clusters of $M_{\star}>10^{4.5}\rm\,M_{\odot}$ and $\sigma_{\rm SFH}<3\rm Myr$ (see definition of $\sigma_{\rm SFH}$ in Section \ref{sec:5.1}). By $z\sim8$, these two metrics diverge significantly: while only $30\text{--}40\%$ of the stellar mass remains in bound clusters, $\sim 60\%$ originated in them. This implies that roughly half of the stars currently found in the diffuse field were actually born in clusters that have since dissolved. The remaining $\sim40\%$ stars formed in NSC, lower-mass clusters ($<10^{4.5}\rm\,M_{\odot}$) or diffuse environments. This cluster birth fraction increases significantly at higher redshifts ($z>15$), indicating that virtually all early star formation occurred in massive clusters. Notably, the very first star particles in our simulation appear at $z=34$ ($84\rm\,Myr$ after the Big Bang) and are exclusively located within a single dense cluster of $4.2\times10^4\rm\,M_\odot$. 

Moreover, if we also include stars in the NSC, the fraction of stars born in clusters rises to $\sim 90\%$ across all redshifts \textbf{(not shown in this figure)}. Consequently, it is conceivable that at sufficiently high resolution, nearly all star formation at cosmic dawn proceeds via the FFB channel, a hypothesis that warrants further investigation in future simulations.


\subsection{Mass Function of Clusters}\label{sec:4.3}

\begin{figure}
    \includegraphics[width=\linewidth]{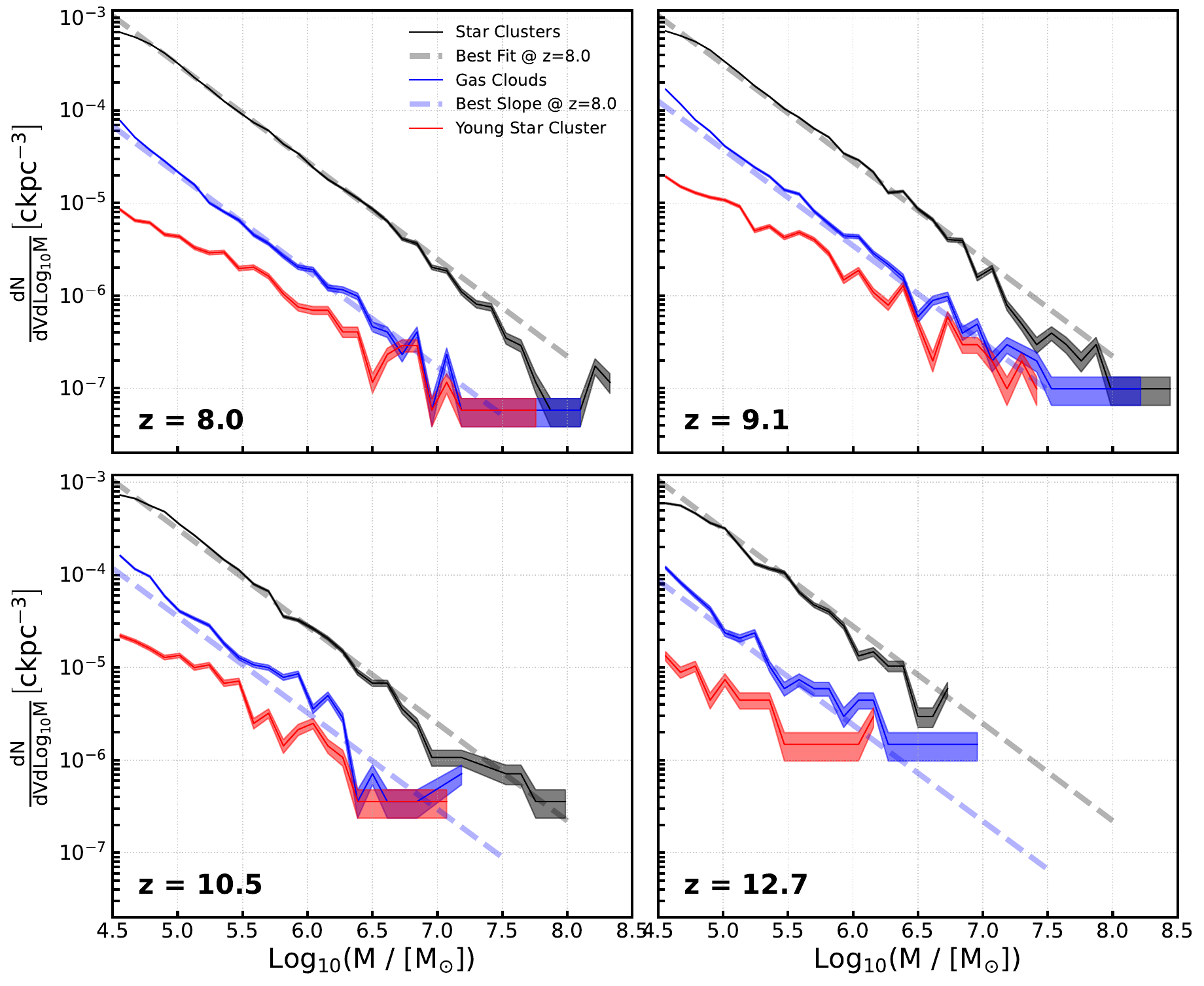}
    \caption{Mass functions within the virial radius. Each panel displays the results at a specific redshift (from upper left to lower right: $z=8.0,\ 9.1,\ 10.5,\ 12.7$). The black, blue, and red curves represent the mass function of star clusters, gas clouds, and young star clusters (defined as $\rm age<5Myr$), respectively. Shaded regions indicate Poisson errors in each mass bin. The gray dashed line shows the best fit (see Equation \ref{equ:massfunc}) to the star cluster mass function at $z\sim8$. The blue dashed lines indicate the best-fit slope ($\alpha=-1.05$) derived from the gas cloud at $z\sim8$, with the normalization adjusted for each redshift.} 
    \label{fig:massfunc}
\end{figure}
 
We now examine the mass function of the identified clusters in four redshifts (i.e., $z=12.7,\ 10.5,\ 9.1,\ 8.0$), which is presented in the four panels of Figure \ref{fig:massfunc}. The three coloured curves represent the clusters number density $\phi(M)$ in unit \emph{comoving volume} and in logarithmic mass bin, e.g.,
\begin{equation}
    \phi(M)=\frac{\mathrm{d}N}{\mathrm{d}V\mathrm{d}\log_{10}M}.
\end{equation} 
The shaded regions represent the corresponding Poisson errors in each mass bin ($\propto \pm n^{1/2}$). As seen, the star clusters follow an approximate power‑law from $\sim10^{4.7}\rm M_\odot$ until a redshift-dependent truncation mass, where the power-law breaks as the massive clusters become rare. Remarkably, the star cluster mass function is rather universal: both the slope and intercept show no prominent evolution across redshifts. The gray dashed line represents the best fit to the power-law of star cluster mass function at $z=8.0$, which is given by:
\begin{equation}
    \phi(M_\star)=\beta \left(\frac{M_{\star}}{\rm M_{\odot}}\right)^\alpha,\quad \alpha=-1.06,\;\beta=61.9\rm\ ckpc^{-3}.\label{equ:massfunc}
\end{equation}
The best-fit slope $\alpha\sim-1$, showing a scale-free mass distribution, that is, equal mass per logarithmic interval. By integrating the mass function (\ref{equ:massfunc}) within a mass range $\left[M_{\star}, M_{\rm max}\right]$, we get the expression of cluster number density:
\begin{equation}
    n\left(>M_\star\right)=\frac{\mathrm{d}N}{\mathrm{d}V}=\frac{\beta \left(M_{\star}^{\alpha}-M_{\rm max}^\alpha\right)}{-\alpha \rm M^{\alpha}_{\odot}\ln10}\simeq\frac{\beta}{-\alpha\ln10}\left(\frac{M_{\star}}{\rm M_{\odot}}\right)^\alpha.
\end{equation}
Specific to our simulation, the total star cluster number massive than $M_{\star}$ within the virial radius can be estimated by:
\begin{equation}
\begin{aligned}
    N\left(>M_{\star}\right)=nV_{\rm vir}\simeq&\frac{\beta M_{\star}^{\alpha}M_{\rm vir}}{-\alpha\ \Delta_{\rm c}\ \rho_{\rm c,0}\ \Omega_m\ \rm M_{\odot}^\alpha\ \ln10}\\
    \simeq&3.2\times10^{-3}\left(\frac{M_\star}{\rm M_{\odot}}\right)^\alpha M_{\rm vir},\label{equ:ncluster}
\end{aligned}
\end{equation}
where $\Delta_{\rm c}\simeq200$ is the halo overdensity, $\rho_{\rm c,0}$ is the cosmic critical density at $z=0$. Moreover, under the approximation of $\alpha=-1$, the total clusters mass density has the following simple form:
\begin{equation}
    \frac1{V_{\rm vir}}\sum_{M_{\rm min}}^{M_{\rm max}}M_{\star}=\beta\log_{10}\left(\frac{M_{\rm max}}{M_{\rm min}}\right).
\end{equation}
and the mean cluster mass is given by:
\begin{equation}
    \left\langle M_\star\right\rangle=M_{\rm min}\ln\left(\frac{M_{\rm max}}{M_{\rm min}}\right).\label{equ:meanmass}
\end{equation}
One easy to see that the parameter $\beta$ represents the star clusters mass density per logarithmic interval. It is interesting to compare $\beta$ with another redshift-independent value, the comoving baryon density within the virial radius, $\Delta_{\rm c}\bar{\rho}_b\simeq1200\rm M_\odot ckpc^{-3}$. Star clusters spanning each order of mass should equally contribute $\beta/\Delta_{\rm c}\bar{\rho}_b\simeq5.2\%$ to the global star formation efficiency $\varepsilon_s$. Clusters spanning two or three mass orders would contribute the majority of the galaxy's star formation rate (see the black dotted line in Figure \ref{fig:infos2}), further indicating that the majority of stars in our simulated high-$z$ galaxy reside in the cluster environment.

The best-fit to the power-law of the dense gas cloud mass function at $z\sim8.0$ is given by:
\begin{equation}
    \phi(M_g)=\beta \left(\frac{M_{g}}{\rm M_{\odot}}\right)^\alpha,\quad \alpha=-1.02,\;\beta=2.62\rm\ ckpc^{-3}.\label{equ:massfunc_gas}
\end{equation}
The slope $\alpha$ also shows a scale-free mass distribution. As proposed by \cite{2022MNRAS.511..859G}, this distribution can be explained by a droplet growth model ($\dot{m}\propto m$) in a turbulent multi-phase medium. Note that the blue dashed lines in Figure \ref{fig:massfunc} follow the same slope $\alpha$ but with adjusted $\beta$ for different redshift. This reflects that, unlike the star clusters, the formation and consumption of gas clouds are regulated by many factors such as feedback, metallicity, inflows, etc.

Comparison of young star clusters and gas clouds at a given redshift reveals the mass‑dependent efficiency of gas‑to‑star conversion. Specifically, the mass function of young star clusters exhibits a shallower slope than that of gas clouds, indicating that more massive gas clouds convert a larger fraction of their mass into stars. In contrast, lower‑mass gas clouds are preferentially disrupted before significant star formation can occur.

\begin{figure}
    \includegraphics[width=\linewidth]{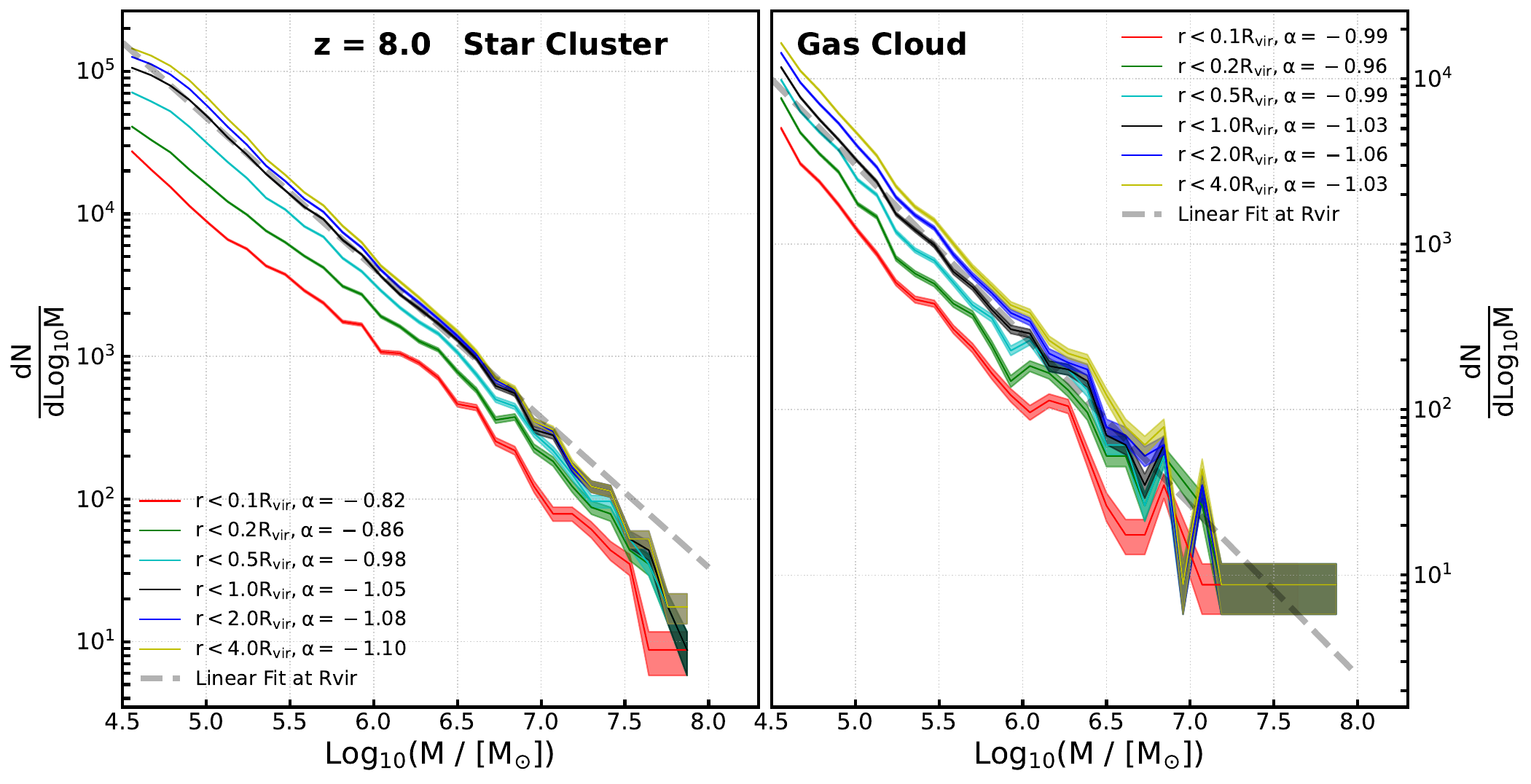}
    \caption{Star cluster (left panel) and gas cloud (right panel) mass distribution at $z\sim8.0$ within different radius. The red, green, cyan, black, blue and yellow curves represent the results within [0.1, 0.2, 0.5, 1.0, 2.0, 4.0]$R_{\rm vir}$. Shaded regions indicate Poisson errors in each mass bin. The gray dashed line shows the best-fit within the virial radius. The best-fit slopes of the linear part of mass distributions are indicated in the legend.}
    \label{fig:massfunc_5r}
\end{figure}

The volume integration of the mass function $\phi(M)$ yields the differential mass distribution $N(M)=\mathrm{d}N/\mathrm{d}\log_{10}M$. Figure \ref{fig:massfunc_5r} illustrates the distinct radial trends in the mass slopes. For gas clouds, the slope remains remarkably consistent at approximately $-1$ across all radii. This universality suggests that gas cloud formation is governed by scale-free gravitational fragmentation and growth throughout the halo. In contrast, the star cluster population exhibits a clear radial gradient: the slope steepens from $-0.82$ in the inner region to $-1.10$ in the outskirts. 

This radial variation in the stellar component is driven by multiple processes. First, as shown in Figure \ref{fig:massfunc}, newly formed star clusters exhibit an intrinsically top-heavy distribution, likely reflecting the rapid ``infant mortality'' of fragile low-mass clusters in the dense high-$z$ environment. Second, tidal stripping converts massive clusters into lower-mass remnants, partially replenishing the low-mass end. Third, subsequent dynamical friction efficiently drives massive clusters toward the galaxy centre, flattening the inner slope while simultaneously depleting the high-mass tail in the outskirts. The infalling systems ultimately merge with the NSC and are removed from the cluster population. Remarkably, despite these complex, mass-dependent processes, the global star cluster population within the virial radius integrates to an approximately scale-free distribution.

\subsection{Composition of star clusters}\label{sec:4.4}

\begin{figure}
    \includegraphics[width=\linewidth]{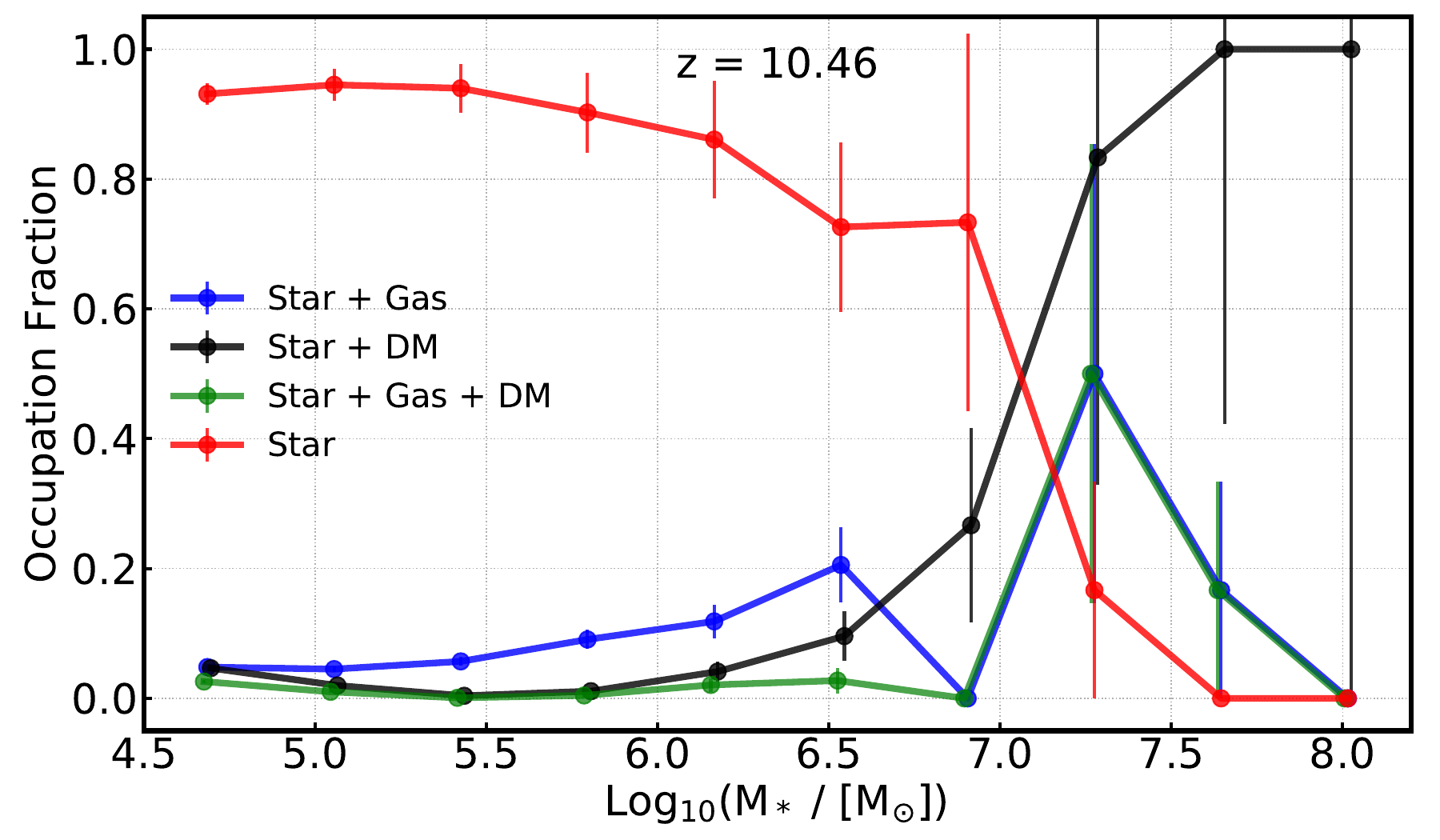}
    \caption{Occupation fractions (see definition in main text) of dark matter and gas in star clusters as a function of mass. The red curve indicates the fraction of pure star clusters, black shows clusters containing dark matter, blue shows those containing gas, and green denotes clusters that contain both dark matter and gas. The Poisson errors are shown for each fraction.}
    \label{fig:occup}
\end{figure}

We further quantify the composition of star clusters. To determine whether a dark matter or gas particle is gravitationally bound to star clusters, using the additional outputs between 445Myr and 460Myr, we trace the evolution of all clusters backward by $1\rm Myr$ and examine whether this particle remains within the cluster (longer tracking do not change the results significantly). In Figure \ref{fig:occup}, we quantify the composition of star clusters at $z=10.46$. For each mass bin, we calculate the following four occupation fractions:
\begin{equation}
    \frac{N_{\rm cluster}\left(\rm s,g\right)}{N_{\rm cluster}},\quad\frac{N_{\rm cluster}\left(\rm s,d\right)}{N_{\rm cluster}},\quad\frac{N_{\rm cluster}\left(\rm s,d,g\right)}{N_{\rm cluster}},\quad\frac{N_{\rm cluster}\left(\rm s\right)}{N_{\rm cluster}},
\end{equation}
where $N_{\rm cluster}$ is the total number of star clusters in that mass bin, and the numerator $N_{\rm cluster}(\cdot)$ counts only those clusters that have bounded dark matter particle (d), gas particle (g), or star particle (s), respectively. Note that the four fractions are not mutually exclusive. This metric thus measures, as a function of cluster mass, the fraction of clusters that contain a certain component. From this figure we demonstrate the following:
\begin{enumerate}
    \item For star clusters with masses below $10^{7}\rm M_{\odot}$, more than $50\%$ are purely stellar (without gas or dark matter), whereas this fraction decreases to zero for clusters more massive than $10^{7.5}\rm M_{\odot}$.
    \item Most of the star clusters (with mass below $10^{8}\rm M_{\odot}$) are devoid of gas. Even in star clusters that initially contain gas, the gas undergoes rapid conversion into stars or blows away by stellar feedback during subsequent evolution (for a more detailed discussion, see Section \ref{sec:5.4}).
    \item Most star clusters more massive than $10^{7.2}\rm M_{\odot}$ contain bounded dark matter (and possibly gas), suggesting that they are likely satellite galaxies rather than FFB clusters.
\end{enumerate}

\section{Star Formation in star clusters}\label{sec:5}
\subsection{Star Formation Histories}\label{sec:5.1}

\begin{figure}
    \includegraphics[width=\linewidth]{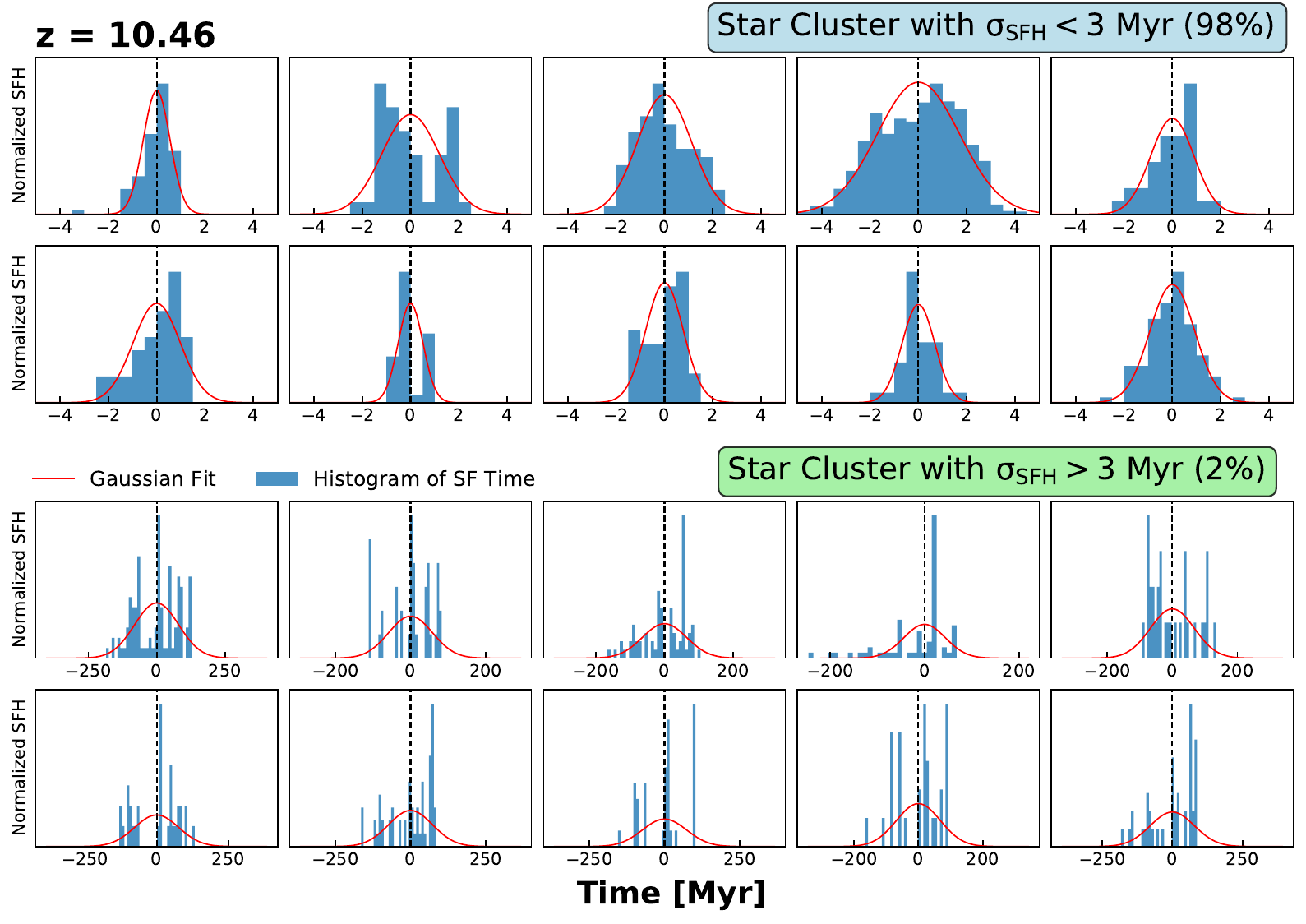}
    \caption{Histograms of normalized SFHs for 20 randomly selected star clusters. The origin of the time axis corresponds to the formation time $t_{\rm form}$ of each cluster, indicated by the vertical dashed line. The time axis is given in units of Myr. The top 10 examples have $\sigma_{\rm SFH}<3\rm Myr$, whereas the bottom 10 examples have $\sigma_{\rm SFH}>3\rm Myr$. Their number fractions are denoted in each bounding box. Red curves show the best-fit Gaussian function of each cluster.}
    \label{fig:sfhist}
\end{figure}

As introduced in Section \ref{sec:1}, a key feature of the FFB scenario is the very short timescale of star formation, enabling a high star formation efficiency before stellar feedback disrupts the starburst. In our simulation, it is straightforward to assess the formation timescale for each identified star cluster. We find that the majority of clusters ($\sim98\%$) are single-population, having experienced only one starburst event. 
We therefore fit the SFH of each cluster with a simple Gaussian distribution, obtaining two parameters: the peak time $\left(t_{\rm form}\right)$ and the standard deviation $\left(\sigma_{\rm SFH}\right)$, which roughly correspond to the formation time and the formation timescale of each cluster, respectively.
Similarly, we fit the metal distribution of each star cluster with a simple Gaussian, with the peak representing the mean metallicity of the cluster $\left(Z_{\rm cluster}\right)$. 

Figure \ref{fig:sfhist} shows the SFH histograms of 20 randomly selected star clusters at $z=10.46$. The top 10 clusters, characterized by $\sigma_{\rm SFH}<3\rm Myr$, exhibit single-population SFHs that are well described by Gaussian functions. In contrast, the bottom 10 clusters, with $\sigma_{\rm SFH}>3\rm Myr$,  typically show multiple-populations or sustained star formations; thus, their SFHs are less well captrued by a single Gaussian function. Nevertheless, the fitted $\sigma_{\rm SFH}$ provides a practical diagnostic parameter to distinguish between single-population and multi-population clusters. 

\subsection{Temporal width of SFH}\label{sec:5.2}

\begin{figure}
    \includegraphics[width=\linewidth]{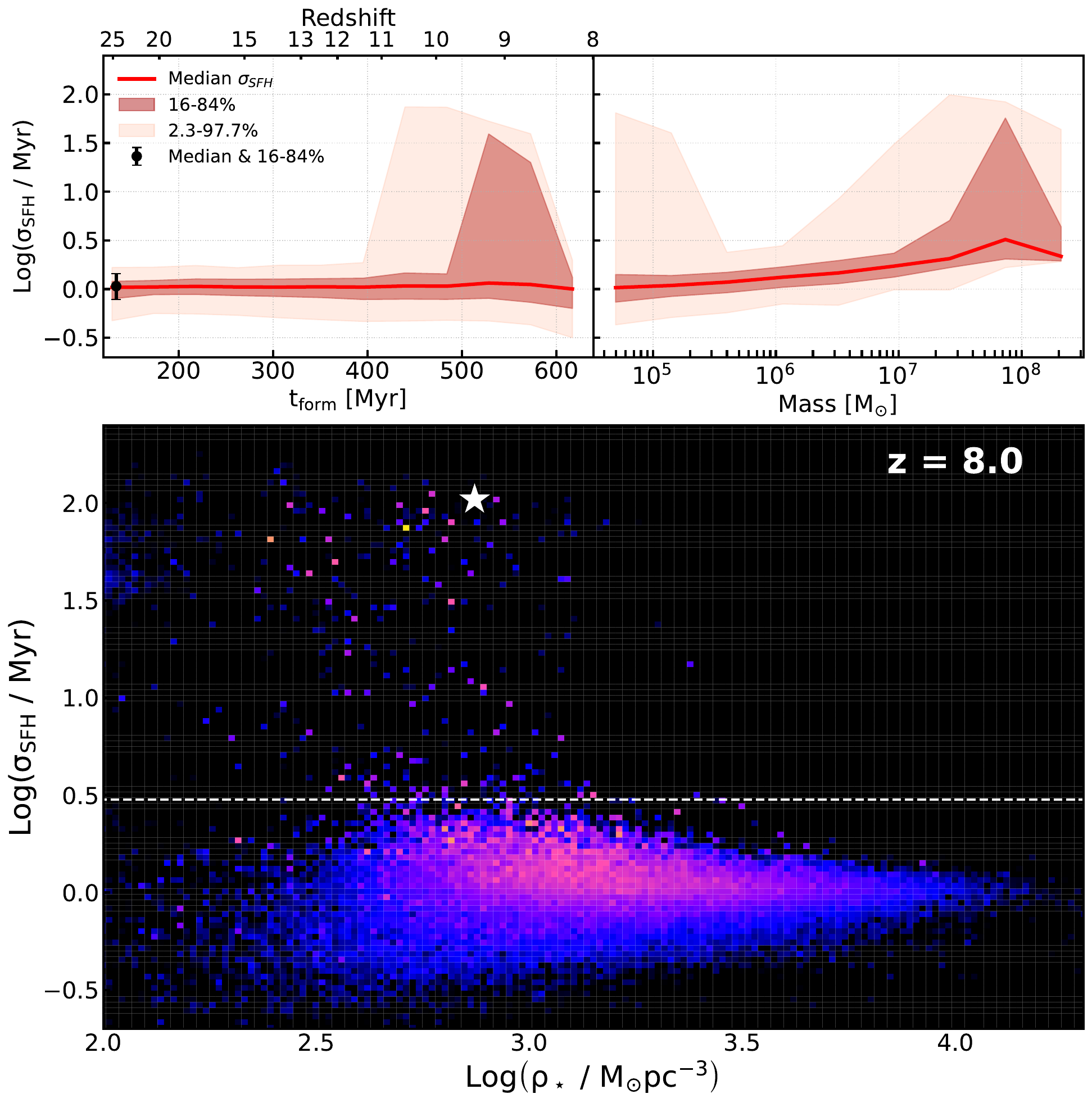}
    \caption{Formation time, cluster mass and stellar density dependence of $\sigma_{\rm SFH}$. The star clusters are identified within $R_{\rm vir}$ at $z\sim8$. The upper left panel shows the SFH standard deviations as functions of their formation time ($t_{\rm form}$, see text for details), while the upper right panel shows the SFH standard deviations as functions of cluster mass. Solid red curves in the two upper panels show the median values, with shaded regions indicating the $16-84$ and $2.3-97.7$ percentiles. The black dot with the error bar shows the median $\sigma_{\rm SFH}$ and $16-84$ percentile. The bottom panel shows 2D histogram distribution of star clusters within $R_{\rm vir}$ in the $\rho-\sigma$ plane. Each pixel's colour represents the total mass of clusters falling within that bin. x-axis is the mean stellar density (see text for definition) and y-axis is the logarithm of $\sigma_{\rm SFH}$. The white dashed line indicates $\sigma_{\rm SFH}=3\rm Myr$. The position of the NSC is marked by the white star. The white arrow roughly shows how cluster moves in this plane when it becomes multi-population.}
    \label{fig:sigma}
\end{figure}

In Figure \ref{fig:sigma} we show the relations between the standard deviation of SFH and three properties of the clusters. These clusters are identified within the virial radius at $z\sim8.0$. The upper left panel shows the distribution of $\sigma_{\rm SFH}$ for clusters formed at different redshifts, about $98\%$ of the star clusters have $\sigma_{\rm SFH}<1.7\rm Myr$, indicating that the majority of star formation in these clusters occurs within a brief window of $2\sigma_{\rm SFH}\sim3.4\rm Myr$. The median $\sigma_{\rm SFH}\sim 1 \rm Myr$ remains nearly constant across all redshifts, except for a noticeable bump around $z\sim9$, where a major merger event triggered the formation of multi-population clusters near the NSC. Notably, this timescale is consistent with the onset delay of SN feedback (Figure \ref{fig:stellarfeedback}), which begins at $3.4\rm\,Myr$ after star formation, confirming that the abundant star clusters in our simulation assemble via the FFB channel.


The upper right panel reveals the mass dependence on the $\sigma_{\rm SFH}$ of clusters. The red curve and the corresponding shaded regions reveal a mild positive relation between $\sigma_{\rm SFH}$ and the mass: more massive star clusters tend to have longer formation timescales. Comparing this with the bottom 10 examples of Figure \ref{fig:sfhist} implies that more massive clusters are more likely to have undergone multiple star formation events or mergers. The bump around $10^{8}\rm M_{\odot}$ corresponds to a population of clusters with sustained star formation. Such massive clusters usually contain dark matter (refer to Figure \ref{fig:occup}), suggesting that they are satellite galaxies. 

In the bottom panel of Figure \ref{fig:sigma}, we further investigate the mass-weighted distribution of star clusters in the $\rho-\sigma$ plane. The x-axis represents the mean stellar density of the clusters, defined as:
\begin{equation}
    \rho_\star=\frac{M_\star}{\frac43\pi abc}
\end{equation}
where $a,b,c$ denote the semi-major, semi-intermediate, and semi-minor axes of the equivalent ellipsoid, respectively. These values are determined by the square roots of the three eigenvalues derived from the inertia tensor of the member stars. This histogram reveals a distinct single-population cluster ``cloud'', which is located below the white dashed line of $3\rm Myr$. This finding further indicates that $\sigma_{\rm SFH}<3\rm Myr$ provides a practical diagnostic criterion to distinguish between single-population and multi-population clusters. The majority of clusters exhibit stellar densities in the range of $10^{2.5-4.0}\rm{M_{\odot}pc^{-3}}$, which is comparable to the central stellar density of the massive host galaxy (Figure \ref{fig:prof}).  

Finally, a small number of star clusters, including the NSC, occupy the region above the dashed line. These clusters usually experienced merger events (see Section \ref{sec:6}) or multiple starburst events. As shown in the upper right panel of this figure, these multi-population clusters are generally more massive than the single-burst clusters found below the white dashed line.


\subsection{Metallicity and Recycling}\label{sec:5.3}

\begin{figure}
    \includegraphics[width=\linewidth]{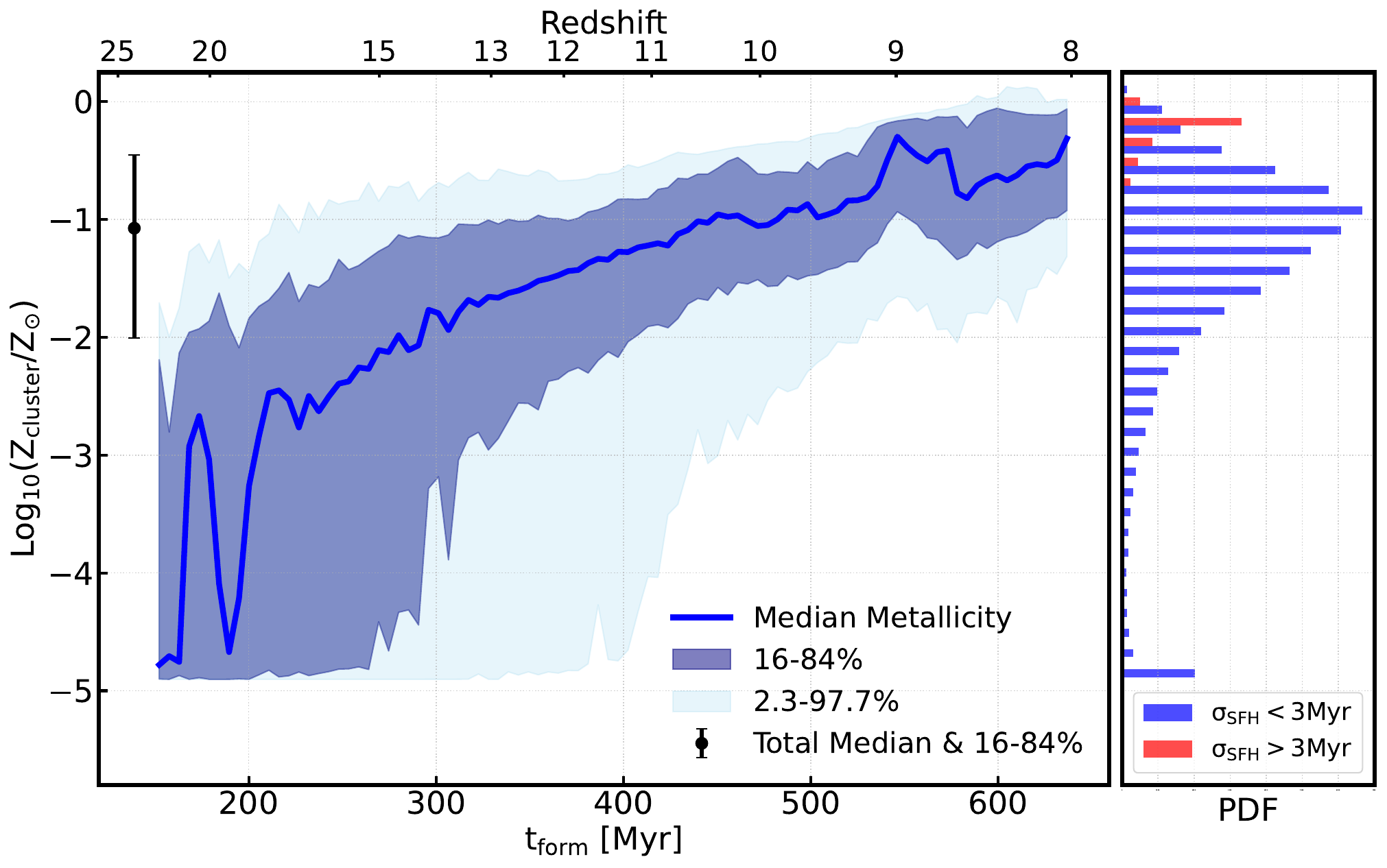}
    \caption{Metallicity evolution and distribution of star clusters within $R_{\rm vir}$. The main panel shows the dependence of cluster metallicity (normalized to $Z_{\odot}$) on formation time, with the solid blue curve representing the median and the shaded regions indicating the $16-84\%$ and $2.3-97.7\%$ percentiles. The black dot associated error bar denote the global median and $16-84$ percentile. The right panel displays the metallicity histograms for single-population ($\sigma_{\rm SFH}<3\rm Myr$) and multi-population ($\sigma_{\rm SFH}>3\rm Myr$) clusters.}
    \label{fig:metals}
\end{figure}

Figure \ref{fig:metals} shows the distribution of the most probable metallicity of star clusters as a function of the formation time. The clusters exhibit a median metallicity of $\log\left(Z/Z_{\odot}\right)=-1.07$, with $68\%$ of them having metallicities in the range $-2.01<\log\left(Z/Z_{\odot}\right)<-0.45$, roughly covering the typical metallicity of the metal-poor GCs in nearby galaxies \citep{1991A&A...247...56F,2024AJ....168...75H}. The blue curve with shaded region shows the metallicity distribution of clusters formed at different redshifts. As seen, the earlier formed clusters are generally more metal-poor, which is a natural consequence of chemical enrichment from accumulated stellar feedback. Gas within the virial radius has already undergone varying degrees of metal enrichment within $200-400\rm Myr$ before being converted to next-generation clusters. The fluctuations of the median metallicity in the early stages reflect this rapid chemical enrichment, which finally leaves an ultra-metal-poor population tail in the metallicity distribution (right panel). In fact, by $z\sim20$, about 20\% of the clusters already have $\log\left(Z/Z_{\odot}\right)>-2$, showing a more rapid baryon recycling compared to the local Universe. 

Unlike local galaxies where supernova ejecta must navigate a diffuse hot halo, the rapid chemical enrichment and baryon recycling at $z\sim10$ is primarily driven by the synergy between intense cold-flow accretion and highly efficient turbulent mixing in the compact interstellar medium. This high-efficiency recycling can be understood qualitatively by comparing the characteristic cooling and free‑fall timescales:
\begin{equation}
\begin{aligned}
    \frac{t_{\rm cool}(z)}{t_{\rm cool}(z=0)}&=\left(\frac{\rho(z)}{\rho(z=0)}\right)^{-1}\simeq\Omega_m^{-1}(1+z)^{-3},\\ \frac{t_{\rm ff}(z)}{t_{\rm ff}(z=0)}&=\left(\frac{\rho(z)}{\rho(z=0)}\right)^{-1/2}\simeq\Omega_m^{-1/2}(1+z)^{-3/2},
\end{aligned}
\end{equation}
which is typically several millions years at $z\sim10$. 
Outflows of $\sim1000\rm km/s$ can only reach a few kpc far, thus inevitably encountering the penetrating cold inflows depicted by Figure \ref{fig:zoom_ffb}. This leads to a ``localized'' baryon recycle, where enriched gas is recaptured and re-incorporated into next-generation clusters within the ISM region, explaining the rapid transition from pristine to enriched states.

As illustrated by the red and blue histograms in the right panel of Figure \ref{fig:metals}, the metallicity distribution of multi-population clusters differs significantly from that of their single-population counterparts. Specifically, multi-population clusters are typically more metal-rich. It is important to emphasize that the star clusters mentioned here are hypothesized as the progenitors of the metal-poor GCs, thus the bimodality observed here is fundamentally distinct from the classic bimodal metallicity distribution of GCs in the local Universe.

\subsection{Local SFE}\label{sec:5.4}

\begin{figure}
    \includegraphics[width=\linewidth]{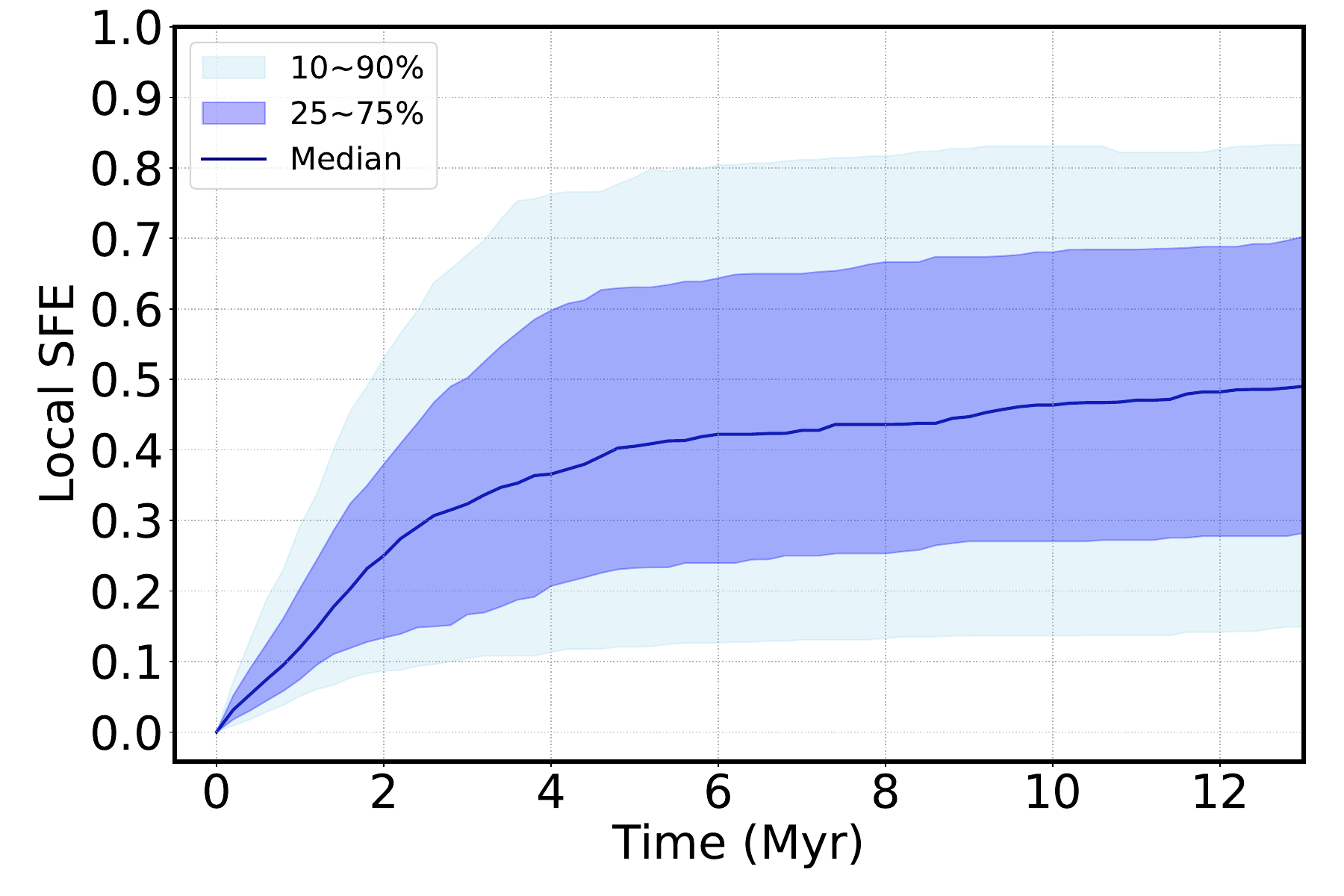}
    \caption{Stacked local SFE of 397 star-forming gas clouds (with mass $>10^{4.5}\rm M_{\odot}$) identified at $z=10.46$. The solid curve represents the median value, while the navy and skyblue band indicates the $25-75\%$ and $10-90\%$ percentiles, respectively.}
    \label{fig:sfe}
\end{figure}

The crucial feature of high-$z$ galaxy formed in the FFB scenario is their relatively high SFE ($\sim0.2$) compared to nearby galaxies, which typically ranges from $10^{-5}$ to $10^{-2}$ depending on the mass of the galaxy. In this subsection, we trace the formation and dynamical evolution of star clusters. To achieve this, as mentioned in Section \ref{sec:2}, we reduce the output interval between 445Myr ($z=10.46$) and 460Myr ($z=10.20$) from 5Myr to 0.2Myr, to resolve the evolution of star formation history in more detail.

We begin by identifying gas clouds with masses greater than $10^{4.5}\rm M_{\odot}$ at 445Myr and tracing their dynamics and composition (gas and star) according to the unique particle ID during the subsequent 15Myr. This procedure enables a direct measurement of the stellar mass fraction in each cluster, which can be interpreted as a ``local SFE''. We find that the local SFE of individual dense gas clouds exhibits step-like increases, with starburst onset times that vary from cluster to cluster. To quantify this behaviour, we select gas clouds that convert at least 4\% of their gas particles into star particles within 1Myr. This criterion yields 397 active clouds ($\sim25\%$ of the sample, as we have shown in Section \ref{sec:4.3}, low mass gas clouds are preferentially disrupted before significant star formation can occur). We then aligned these clouds by their individual starburst initiation times and stacked their ensuing local SFEs. As shown in Figure \ref{fig:sfe}, the stacked profile clearly demonstrates a rapid rise in local SFE from 0 to $\sim0.5\pm0.2$, followed by a plateau of approximately $4\rm Myr$, which is in good agreement with our previous analysis of $\sigma_{\rm SFH}$. Notably, \cite{2017ApJ...834...69L} also found that feedback from young stars extinguishes star formation in dense GMCs within $4\rm\ Myrs$, consistent with the observed age spread of young star clusters (see their Figure 11).

\section{Evolution of star clusters}\label{sec:6}


Having quantified the initial high SFEs of the FFB process, we now shift our focus to the subsequent dynamical fates of these nascent clusters. To determine how many of them survive versus those that contribute to the galaxy's central or diffuse stellar components, we track their dynamical evolutions. To begin, we establish the progenitor–descendant link between the clusters identified in two close snapshots. We define the match ratio between clusters as follows:
\begin{equation}
    R(A,B)=\frac{c^2}{ab}\label{equ:matchratio}
\end{equation}
where cluster $A$ in snapshot $N$ contains $a$ stellar particles, cluster $B$ in snapshot $(N+1)$ contains $b$ particles, and $c$ is the number of shared particles (based on particle IDs) between the two clusters. A given cluster in one snapshot may be linked to multiple clusters in the next snapshot, each with a different match ratio. We assign the cluster with the highest match ratio as the unique descendant of cluster $A$. In contrast, the cluster $B$ in snapshot $(N+1)$ may have multiple progenitors in snapshot $N$, which typically indicates a merging event. By tracking these connections, we broadly classify the evolutionary pathways of star clusters into the following 6 categories:
\begin{enumerate}
    \item \textbf{Mass-growing clusters}: These are systems undergoing gradual mass accretion, defined as a subsequent mass increase exceeding 10\% relative to their mass at the start of the tracking period.
    \item \textbf{Cluster mergers}: This category includes clusters that coalesce with other star clusters (excluding the NSC) during the tracking period.
    \item \textbf{Accretion by the NSC}: Clusters that are captured and incorporated into the nuclear star cluster during the tracking period.
    \item \textbf{Stripped clusters}: These systems exhibit a gradual mass loss of more than 10\%, typically driven by tidal stripping or other external dynamical processes.
    \item \textbf{Stable clusters}: Clusters that maintain a nearly constant mass, with fluctuations remaining within ±10\% of their initial values.
    \item \textbf{Dissipated clusters}: This category comprises clusters whose descendants can no longer be identified as bound structures above our resolution threshold. This indicates complete tidal disruption or evaporation, though it is worth noting that the dispersed stars from these clusters may eventually be captured by the NSC and subsequently reclassified under category (iii).
\end{enumerate}

\begin{figure}
    \includegraphics[width=\linewidth]{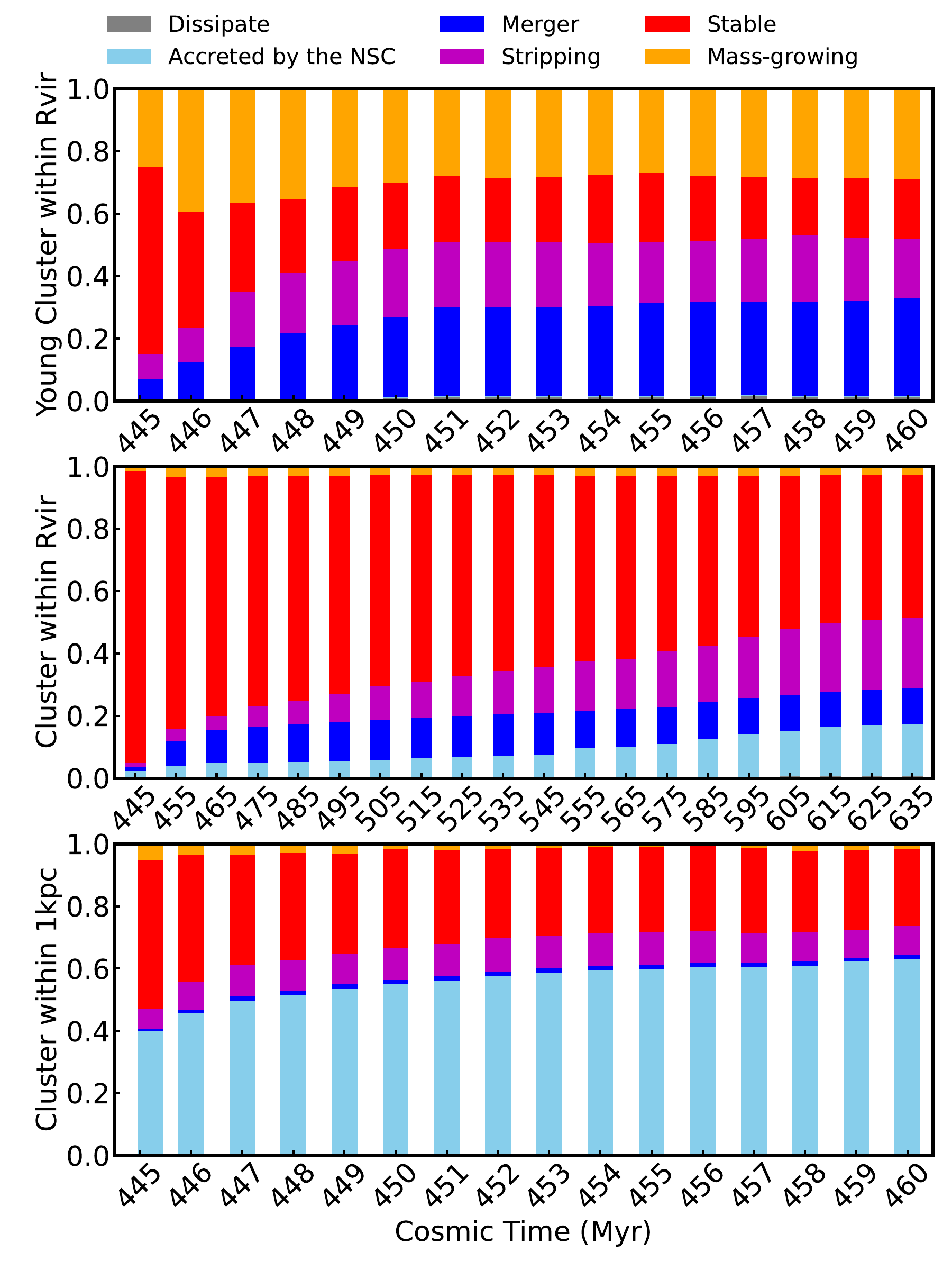}
    \caption{Evolution of star clusters identified at $z=10.46$ in subsequent snapshots. From top to bottom, the panels show subsets of clusters that: (1) young age ($<5\rm Myr$) and reside within $R_{\rm vir}$; (2) reside within $R_{\rm vir}$; (3) reside within $1\rm kpc$. The time intervals between bars are $10\rm Myr$ for the middle panel and $1\rm Myr$ for the other two panels. The colours represent the fraction of clusters in different evolutionary states: dissipation (gray), accretion by the NSC (skyblue), merger with other clusters (blue), stripping (magenta), stable (red) and smooth growth (orange). See the main text for details on the criteria used to define these evolutionary states.}
    \label{fig:frac}
\end{figure}

Figure \ref{fig:frac} presents the evolutionary results of star clusters identified at $z=10.46$. The three panels correspond to cluster subsets selected by their spatial location and formation epoch. Each bar denotes the fraction of clusters in various evolutionary states in subsequent snapshots.

The top panel investigates the evolution of young clusters (aged $<5\rm Myr$) . It shows that about 30\% of the young clusters undergo merger events within $6\rm Myr$ of their formation. Other clusters follow diverse evolutionary trajectories such as tidal stripping, further mass growth. The increasing merger fraction is driven by the structure of the filamentary gas clouds (see Figure \ref{fig:sgclusters}), which typically fragment to form several clusters along the longitudinal axis of the filament during gravitational collapse. Due to their spatial proximity and shared kinematics, these clusters then rapidly coalesce within a short timescale ($<6\rm Myr$) after their birth.

For clusters within the virial radius (middle panel), a growing proportion of the population undergoes tidal stripping over time. There is also a non-negligible contribution from mergers and a rising fraction of clusters accreted by the NSC. This result reinforces the conclusion that a portion of the diffuse stellar component originates from stars stripped from clusters formed in the FFB scenario (see also the black and red curves in Figure \ref{fig:sfrfrac}). 

The bottom panel shows that, within a few tens of millions of years, about 80\% of the clusters located within the central $1\rm kpc$ are accreted by the NSC (skyblue bars). A small fraction undergoes tidal stripping or remain stable, while other evolutionary pathways are negligible. This indicates that the innermost region is highly dynamic and gravitationally dominated by the NSC, leading to cluster disruption in very short timescales.



Combining the insights from these three panels, a coherent and environment-dependent lifecycle for high-$z$ star clusters emerges. Upon formation, young clusters frequently undergo rapid mass growth and major mergers ($<6\rm\,Myr$) driven by the dense, filamentary geometry of their natal gas. Following this initial assembly, their fates strongly diverge based on their spatial location. Clusters in the outer halo typically enter a prolonged phase of gradual tidal stripping or remain structurally stable, continuously feeding the diffuse stellar field. Conversely, clusters residing in or migrating to the central regions undergo rapid orbital decay, culminating in accretion by the NSC within a few tens of millions of years. Crucially, this direct evolutionary tracking validates the dynamical scenario inferred from the cluster mass functions: the intense central tidal field and the NSC act as an efficient destruction sink, profoundly reshaping the surviving cluster population and supplying the bulk of the galaxy's diffuse stars.

\section{Shape of star clusters}\label{sec:7}

\begin{figure}
    \includegraphics[width=\linewidth]{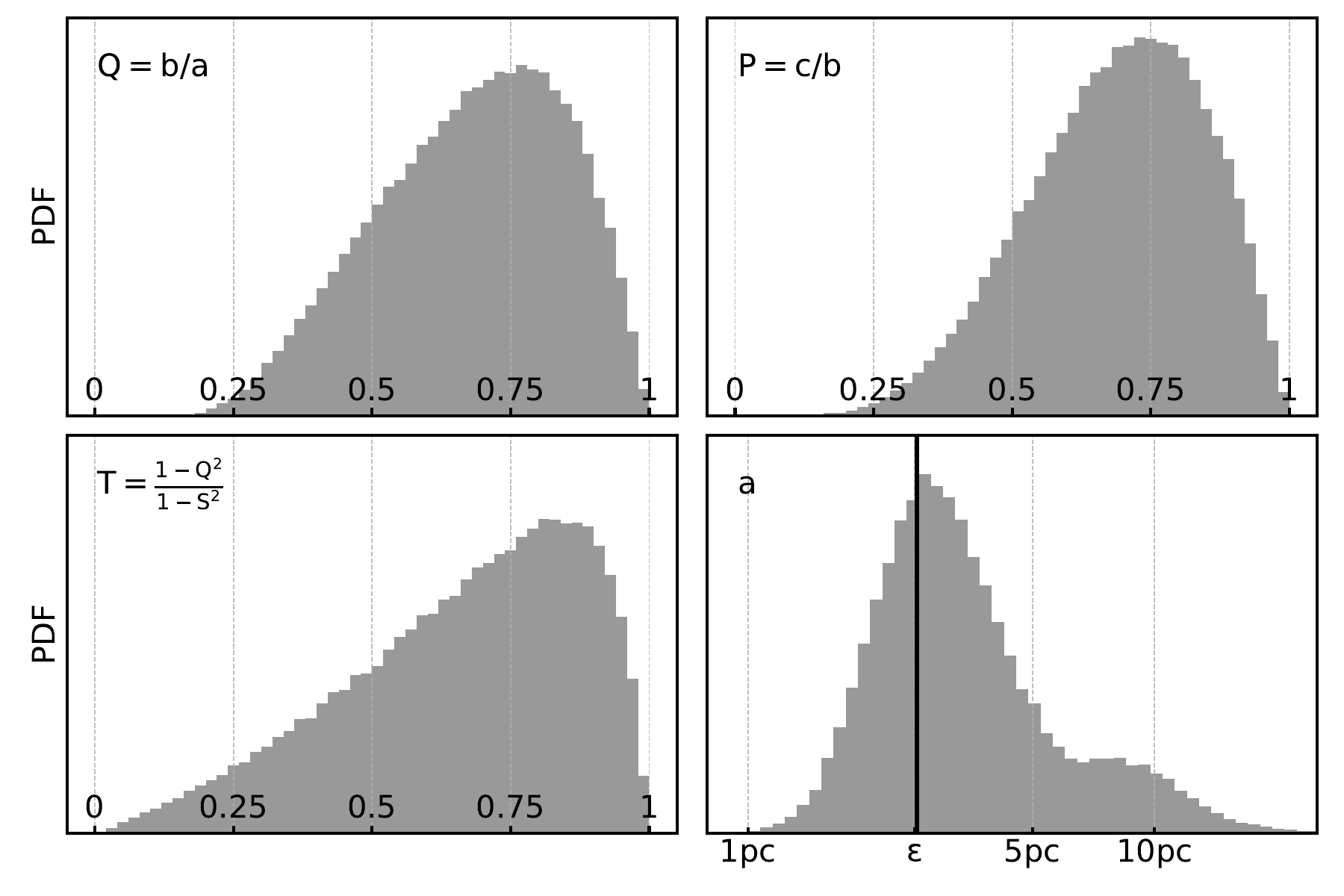}
    \caption{The distribution of three shape parameters ($Q,\ P,\ T$) of star clusters within $R_{\rm vir}$ at $z\sim8.0$. The lower right panel shows the distribution of the semi-major axes of clusters (in physical parsec). For comparison, the softening length of star particle is indicated in this panel.}
    \label{fig:shape}
\end{figure}

In section \ref{sec:5.2} we calculate the semi-major ($a$), semi-intermediate ($b$), and semi-minor axes ($c$) of the equivalent ellipsoid for each star cluster. These allow for a convenient analysis of their shapes. The ellipticity of an ellipsoid can be characterized by three axial ratios (only two of them are independent), e.g.,
\begin{equation}
    S=\frac{c}{a}\,,\quad P=\frac{c}{b}\,,\quad Q=\frac{b}{a}\,.\label{equ:shape1}
\end{equation}
The cluster is oblate if $Q<P$ and it is prolate if $Q>P$. When $P\sim Q$ the cluster is triaxial. We follow \cite{1991ApJ...383..112F} and use 
\begin{equation}
    T=\frac{1-Q^2}{1-S^2}=\frac{a^2-b^2}{a^2-c^2},\label{equ:shape2}
\end{equation}
to qualify the triaxiality of clusters. Both parameters range from 0 to 1. The four upper panels of Figure \ref{fig:shape} present the distribution of three shape parameters, as well as the physical scale of the semi-major axes. As seen, three histograms peak at $P\sim0.75$, $Q\sim0.76$, and $T\sim0.8$, indicating that the star clusters are highly triaxial in shape. Cluster physical sizes, characterized by their semi-major axes, lie between $1\sim10\rm pc$ range, which can be further classified into small ($\sim3\rm pc$) and large ($\sim10\rm pc$) populations. Note that for clusters with sizes comparable to or smaller than the softening length, our simulation cannot accurately resolve their internal dynamic evolutions. 

\begin{figure}
    \includegraphics[width=\linewidth]{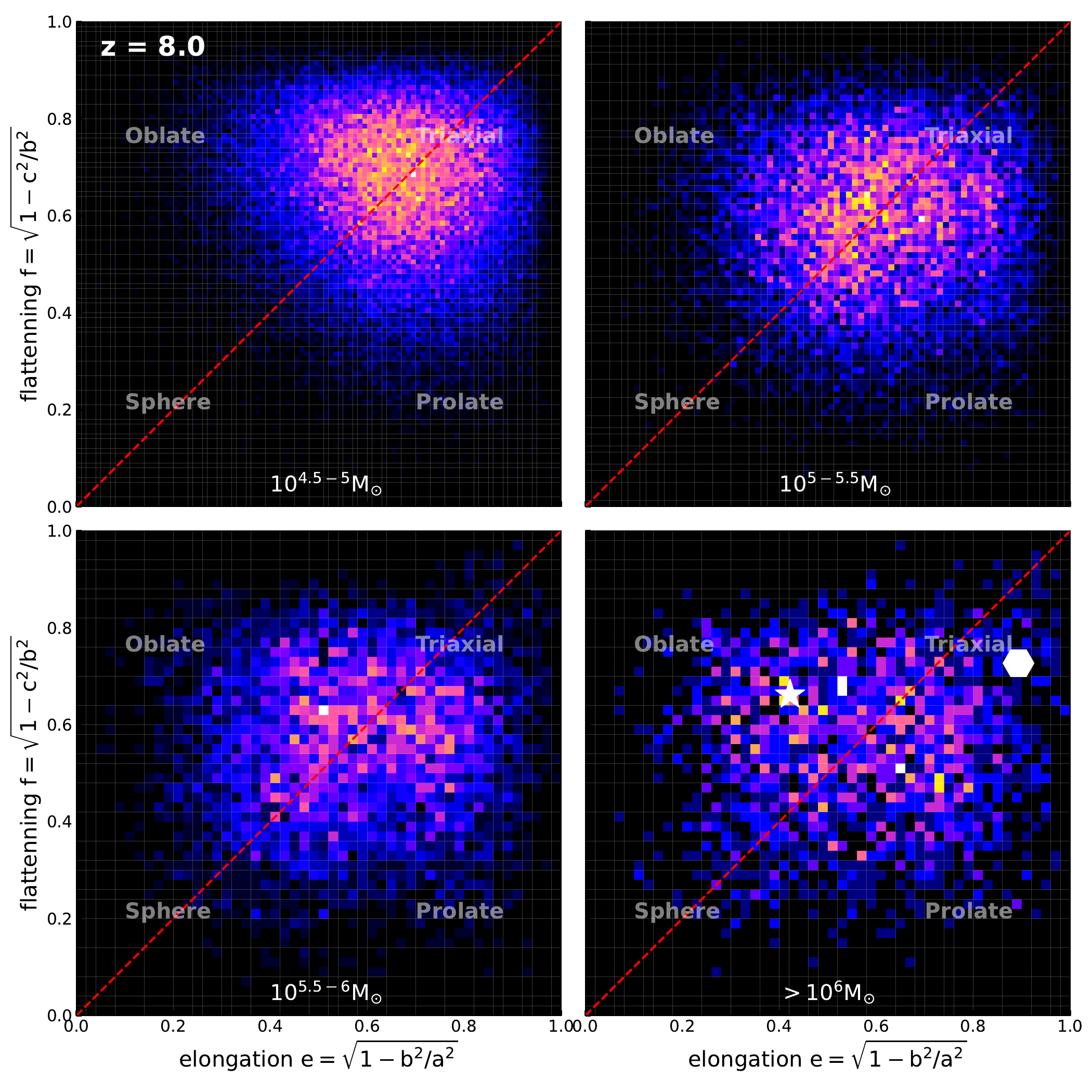}
    \caption{Distribution of star clusters within $4R_{\rm vir}$ in the $e-f$ plane. The red dashed line denotes pure triaxial systems, ranging from spherical to very elongated triaxial configurations. Note that the triaxiality here is a little different with Equation \ref{equ:shape2}. The four panels represent clusters in different mass range, as indicated at the bottom of each panel. The position of the NSC is marked by the white star. The white hexagon represents the global shape parameters derived from the spatial distribution of all clusters (weighted by mass).}
    \label{fig:ef}
\end{figure}

For the purpose of a more comprehensive shape analysis, we follow \cite{2016MNRAS.458.4477T} and further define the parameter of elongation ($e$) and flattening ($f$):
\begin{equation}
    e=\sqrt{1-Q^2},\quad f=\sqrt{1-P^2},\label{shape3}
\end{equation}
which also range from 0 to 1. The $e-f$ plane provides a robust framework for investigating the morphological distribution of the star clusters. Since the four primary geometric configurations: spherical, oblate, prolate, and triaxial, occupy distinct regions in this plane, the parameters of elongation ($e$) and flattening ($f$) characterize the cluster shapes with high precision. Figure \ref{fig:ef} presents 2D histograms of these parameters, subdivided into four mass bins arranged from the top-left to the bottom-right panels. From this analysis, we derive the following four insights regarding cluster shapes:
\begin{enumerate}
    \item Clusters are predominantly non-spherical across all mass ranges, showing no significant statistical preference for either oblate or prolate geometries.
    \item Clusters with masses below $10^5\rm M_{\odot}$ exhibit a higher degree of triaxiality compared to more massive populations.
    \item The NSC (indicated by the white star in the lower-right panel) is distinctly oblate, indicating a clear rotational support signature.
    \item The global shape parameter of the entire cluster system, derived from the spatial distribution of all clusters (indicated by the white hexagon), is highly triaxial and prolate. The semi-major axis of this distribution aligns with the orientation of the filamentary cold gas inflows, as seen in the stellar distribution panel of Figure \ref{fig:zoom_ffb}). 
\end{enumerate}

\section{Comparison to Nearby Globular Clusters}\label{sec:8}


The star clusters resolved in our simulation exhibit a striking physical resemblance to the metal-poor GCs observed in the local Universe. Their mass range of $10^{4.5-6.5}\rm M_\odot$ fully encompasses that of nearby GCs \citep[e.g.,][]{2018MNRAS.478.1520B}, and their inner stellar densities are comparable, implying similar physical scales (see the lower right panel of Figure \ref{fig:shape}). Crucially, over 80\% of the clusters below $10^{6.5}\rm M_\odot$ are found to be purely stellar and devoid of gas, mirroring the gas-poor nature of observed GCs. These clusters are overwhelmingly single-population systems, characterized by feedback-free starbursts with star formation histories well-fitted by a narrow Gaussian with $\sigma_{\rm SFH}\sim1.1\rm Myr$. Furthermore, their metallicity distribution spanning $-2.01<\log\left(Z/Z_{\odot}\right)<-0.45$ aligns closely with the metal-poor GC populations identified in local galaxy surveys.

Beyond individual properties, the global statistics of these clusters provide a compelling link to local galaxy clusters like Virgo and Coma. Surveys of the Virgo cluster \citep{2014ApJ...794..103D} estimate a total GC population of $N=67,300\pm14,400$. \cite{2011ApJ...730...23P} constrain the Coma cluster to contain $47,000\pm1600$ GCs within a central radius of $520\rm kpc$, suggesting a total population that may exceed $10^5$. Our simulation predicts the formation of approximately $N\sim5\times10^4$ star clusters more massive than $10^{4.5}\rm M_{\odot}$ by the end of the FFB phase (Equation (\ref{equ:ncluster})), a count that is broadly consistent with these local censuses when subsequent hierarchical mergers are considered. Combining Equations (\ref{equ:ncluster}) and (\ref{equ:meanmass}) under the assumption of $\alpha=-1$, the cluster mass fraction $\eta$ in our simulated system can be estimated as:
\begin{equation}
    \eta=\frac{\left\langle M_{\star}\right\rangle N\left(>M_{\rm min}\right)}{M_{\rm vir}}\simeq\frac{\beta}{\Delta_{\rm c}\rho_{\rm c,0}\Omega_m}\log_{10}\left(\frac{M_{\rm max}}{M_{\rm min}}\right),
\end{equation}
which is approximately 1-2\% provided that the dynamic range of cluster masses does not vary significantly between systems. As the system’s
virial mass scales up from $10^{12}\rm M_{\odot}$ at cosmic dawn to $10^{15}\rm M_{\odot}$ today, the mass fraction will naturally decrease to match the empirical value of $\eta\sim2.9\times10^{-5}$ reported in local galaxies \citep{2017ApJ...836...67H}.

It is important to address the discrepancy between the predominantly single-population nature of our simulated clusters and the multiple-populations (MPs) ubiquitous in Galactic GCs. First, we clarify that the ``single-population'' designation in this work is a temporal classification rather than variations in chemical abundances. It refers to the rapid assembly of cluster mass within a narrow formation window ($\sigma_{\rm SFH}<3\rm Myr$). This synchronized birth is a direct physical consequence of the fiducial SNe feedback model implemented in \texttt{GIZMO}, which is calibrated to match the energy injection rates predicted by {\sc starburst99} (see Figure \ref{fig:stellarfeedback}).

On the other hand, while our model reproduces the coeval ``single-burst'' formation characteristic of GCs within a narrow time window, we indeed observe a diversity in chemical abundances, as well as a positive C-O correlation within the resulting star clusters. However, because the chemical yields table in \texttt{GIZMO} does not explicitly account for specific stellar evolutionary pathways, such as Fast-Rotating Massive Stars (FRMS) or Supermassive Stars (SMS), the simulation does not reproduce the C-N anti-correlation typically driven by the CNO cycle—a classic signature of MPs in observed GCs. This discrepancy in detailed abundance patterns, along with a broad metallicity dispersion, also emerges in other high-resolution hydro-simulations. As highlighted in \cite{2016ApJ...823...52K}, such a wide dispersion is likely a natural consequence of rapid, inhomogeneous self-enrichment: a significant fraction of stars form from relatively pristine gas before the first SNe emerge to pollute the natal gas clouds, while subsequent SNe ejecta are distributed non-uniformly throughout the star cluster. Furthermore, our current simulation does not fully capture such small-scale turbulent metal mixing, which might otherwise smooth out these chemical gradients. The inclusion of more sophisticated internal mixing and recycling models, along with comprehensive yield tables incorporating FRMS and SMS, will be necessary in future work to reconcile the temporal ``single-burst'' nature of high-redshift star clusters with the detailed chemical homogeneity and abundance patterns observed in their local survivors.

Indeed, bridging these high-$z$ progenitors with present-day GCs requires accounting for $13\rm\ Gyr$ of complex evolution. Over cosmic time, processes such as two-body relaxation and tidal stripping are expected to reshape these clusters, rendering them more spherical and bringing their density profiles into closer agreement with King models \citep{1962AJ.....67..471K}. Although many clusters in the inner regions are rapidly accreted by the NSC or stripped to form the diffuse stellar halo, the resilient surviving population constitutes the metal-poor GCs we observe today. Given the ubiquitous presence of GCs in the local Universe, the FFB channel is likely a universal mechanism prevalent across a wide range of mass scales, a hypothesis that warrants further investigation in future simulations of different mass regimes.

\section{Conclusion}\label{sec:9}

\begin{table*}
    \centering
    \begin{tabular}{c|c|c}
       \hline\hline
        & FFB Prediction & This Work \\ \hline
        Filamentary Cold Gas Penetrate & yes & yes \\ \hline
        $\varepsilon_s$ & $>0.1$  & $0.2-0.3$ \\ \hline
        $t_{\rm ff}$ & $\sim1\rm Myr$ & $\sigma_{\rm SFH}<3\rm Myr$ \\ \hline
        $\Sigma_{\rm gas}$ & $>3\times10^3\rm M_{\odot}pc^{-2}$ & $>3\times10^3\rm M_{\odot}pc^{-2}$ \\ \hline
        $\rho_{\star,\rm clump}$ & $>10^2\rm M_{\odot}pc^{-3}$ ($n_{\rm gas}>3000\rm cm^{-3}$) & $10^{2.5-4.0}\rm M_{\odot}pc^{-3}$ \\ \hline
        $Z/Z_{\odot}$ & $<0.2$ & 0.085(med), $0.0097\sim0.35$  \\ \hline
        Star Formation Rate & $65\rm M_{\odot}yr^{-1}$ & $\sim100\rm M_{\odot}yr^{-1}$ \\ \hline
        Star Formation History & bursty, $\sim 10{-}20\rm Myr$ period & bursty, $\sim 10{-}20\rm Myr$ period \\ \hline
        $N_{\rm cluster}\left(\ge10^6\rm M_{\odot}\right)$ & $\sim10^3$ & $\sim300$ at $z=10$ \\ \hline
        Speed of Outflows & $727\sim3333\rm km/s$ & $\sim2000\rm km/s$ \\ \hline
         \hline
    \end{tabular}
    \caption{Comparison between theoretical predictions for typical FFB galaxies at $z=10$ and simulation results of this work.}
    \label{tab:ffbtable}
\end{table*} 

In this work, we utilized an ultra-high-resolution cosmological zoom-in hydrodynamical simulation to investigate the formation of a massive galaxy ($M_{\rm vir} \simeq 1.8\times10^{11}\rm\,M_{\odot}$ at $z \simeq 10.46$). By implementing a 3.4 Myr delay in supernova feedback, we explicitly resolved the feedback-free starburst (FFB) process in dense gas clouds down to $10^{4.5}\rm\,M_\odot$. Our results provide a robust, first-principles theoretical counterpart to the ultra-luminous, compact high-$z$ galaxies recently unveiled by JWST. As summarized in Table \ref{tab:ffbtable}, our hydrodynamical findings comprehensively validate the analytical predictions of the FFB model (D23 and L24). The primary physical insights of this study are as follows:

\begin{itemize}
     \item \textbf{Validation of the FFB Mechanism in Extreme Environments:} The simulated galaxy core (the NSC) is extraordinarily compact ($R_{\rm e} \sim 1$ kpc) and dense ($\Sigma_* > 10^5\rm\,M_{\odot} pc^{-2}$ within the inner 100 pc). Fuelled by deeply penetrating cold filamentary inflows, the system achieves a highly efficient global star formation phase (SFE $\sim 0.2-0.3$). During this phase, the rapid stellar mass assembly at the galaxy center completely outpaces dark matter and gas accumulation, producing a strongly star-dominated nucleus.
    
     \item \textbf{A Cluster-Dominated Era of Star Formation:} We find that at cosmic dawn, star formation is overwhelmingly clustered. We identified over $10^5$ individual star clusters (following a scale-free mass function with $\alpha \simeq -1.06$), which account for approximately 90\% of all stars formed in this epoch, \textbf{and at a given time constitute $30\text{--}40\%$ of the total stellar mass}. These clusters assemble via rapid, feedback-free bursts ($\sigma_{\rm SFH} < 3\rm\,Myr$), achieving high local efficiencies ($\sim 0.5$) and evolving into gas-poor, strongly bound, and metal-poor clusters before supernova feedback can disrupt their natal gas clouds.
    
     \item \textbf{Divergent Dynamical Fates of Clusters:} Once formed, the survival and evolution of these clusters are strictly dictated by their spatial environments. In the highly dynamic central 1 kpc, clusters undergo rapid orbital decay and merge to assemble the massive NSC. Conversely, clusters in the extended halo are dominated by gradual tidal stripping. This environment-dependent filtering dictates that surviving massive clusters segregate to the centre, while disrupted low-mass clusters continuously feed the galaxy's diffuse stellar halo.
    
     \item \textbf{Progenitors of Present-Day GCs:} Although the diversity in light-element abundances observed in local GCs fails to reproduced in our current framework, the surviving clusters in our simulation naturally reproduce many physical properties of metal-poor GCs observed in the local Universe, such as the stellar density ($\rho_{\star}\sim10^{2.5-4}\rm M_{\odot}pc^{-3}$), cluster radius ($a\sim1\text{--}10\rm pc$), gas-poor nature (mostly without gas and dark matter for mass below $10^{7}\rm M_{\odot}$). Furthermore, our derived high-$z$ cluster abundance and initial mass fraction ($\sim 1\text{--}2\%$) broadly consistent with the present-day GC specific frequencies in local galaxy clusters like Virgo and Coma, provided the host halo undergoes significant subsequent dark matter growth. This strongly supports the hypothesis that local GC systems are the relic remnants of FFB star formation at cosmic dawn, having survived $13\rm\ Gyr$ of dynamical reshaping.
\end{itemize}

\section*{Acknowledgements}

This work is dedicated to the memory of our colleague and co-author, Prof. Avishai Dekel, who passed away during the preparation of this manuscript. His initial ideas for the FFB model and his constant encouragement were pivotal to this research; indeed, this article would not have been possible without him. More broadly, we would not be who we are today without him. We are deeply grateful to have known and worked with him.

This work is supported by the National Key Research and Development Program of China (No.2022YFA1602903, No. 2023YFB3002502), the National Natural Science Foundation of China (NFSC) (No. 12595314, 12533007, 12233005, 12547104). HZC acknowledges the cosmology simulation database (CSD) in the National Basic Science Data Center (NBSDC) and its funds the NBSDC-DB-10. 
ZZL acknowledges the Marie Skłodowska-Curie Actions Fellowship under the Horizon Europe programme (101109759, ``CuspCore''). AD, ZL and ZY have been supported by the US National Science Foundation (NSF) - US-Israel Binational Science Foundation grants 2023723 and 2023730, and by Israel Science Foundation grant 861/20. We also thank Kartick Chandra Sarkar for kindly providing the {\sc starburst99} results. The simulations and analysis presented in this article were carried out on the SilkRiver Supercomputer of Zhejiang University, located at the Zhejiang University Information Center.

\section*{Data Availability}

The simulation data underlying this article will be shared on reasonable request to the corresponding author.

\bibliographystyle{mnras}
\bibliography{FFB}


\appendix

\bsp	
\label{lastpage}
\end{document}